\documentclass[a4paper,nofootinbib]{revtex4}
\usepackage{graphicx,calc,epsfig}
\usepackage{amsmath,amssymb,amsfonts}

\def\ket#1{| #1 \rangle}

\def\bb1{\textup{\small{1}} \kern-3.6pt \textup{1}\ }

\DeclareFontFamily{U}{rsfs}{}         
\DeclareFontShape{U}{rsfs}{m}{n}{<5> rsfs5 <6><7> rsfs7          %
  <8><9><10><10.95><12><14.4><17.28><20.74><24.88> rsfs10}{}     %
\DeclareMathAlphabet{\mathfs}{U}{rsfs}{m}{n}                     %
\newcommand{\mfs}[1]{\mathfs {#1}}                               %

\newcommand{\sH}{{\mfs H}}

\usepackage{bbm}
\usepackage{fancyhdr}

\newcommand{\ssp}{< {\rm P}^{\va (m)} s,s^{\prime} >}

\newcommand{\Z}{\mathbb{Z}}

\newcommand{\va}{\scriptscriptstyle}

\newcommand{\R}{\mathbb{R}}

\newcommand{\Hp}{{\sH}_{phys}}
\newcommand{\Hk}{{\sH}_{kin}}

\newcommand{\be}{\nopagebreak[3]\begin{equation}}
\newcommand{\ee}{\end{equation}}
\newcommand{\ba}{\nopagebreak[3]\begin{eqnarray}}
\newcommand{\ea}{\end{eqnarray}}

\begin{document}

\title{On the regularization ambiguities in loop quantum gravity}
\author{Alejandro Perez}
\email{perez@cpt.univ-mrs.fr}
\address{Centre de Physique Th\'eorique\footnote{Unit\'e Mixte de Recherche (UMR 6207) du CNRS et des Universit\'es Aix-Marseille I, Aix-Marseille II, et du Sud Toulon-Var; laboratoire afili\'e \`a la FRUMAM (FR 2291).}, Campus de Luminy, 13288 Marseille, France}

\vspace{.5cm}

\begin{abstract}

One of the main achievements of loop quantum gravity is the consistent
quantization of the analog of the Wheeler-DeWitt equation which is
free of ultra-violet divergences. However, ambiguities associated to
the intermediate regularization procedure lead to an apparently
infinite set of possible theories.  The absence of
an UV problem---the existence of well behaved regularization of the
constraints---is intimately linked with the ambiguities arising in
the quantum theory. Among these ambiguities there is the one associated
to the $SU(2)$ unitary representation used in the diffeomorphism
covariant ``point-splitting'' regularization of the non linear
functionals of the connection. This ambiguity is labelled by a
half-integer $m$ and, here, it is referred to as the {\em
$m$-ambiguity}. The aim of this paper is to investigate the important
implications of this ambiguity.

We first study 2+1 gravity (and more generally BF theory) quantized in
the canonical formulation of loop quantum gravity.  Only when the
regularization of the quantum constraints is performed in terms of the
fundamental representation of the gauge group one obtains the usual
topological quantum field theory as a result.  In all other cases
unphysical local degrees of freedom arise at the level of the
regulated theory that conspire against the existence of the continuum
limit. This shows that there is a clear cut choice in the quantization
of the constraints in 2+1 loop quantum gravity.

We then analyze the effects of the ambiguity in 3+1 gravity exhibiting
the existence of spurious solutions for higher representation
quantizations of the Hamiltonian constraint.  Although the analysis is
not complete in 3+1 dimensions---due to the difficulties associated to
the definition of the physical inner product---it provides evidence
supporting the definitions quantum dynamics of loop quantum gravity in
terms of the fundamental representation of the gauge group as the only
consistent possibilities. If the gauge group is $SO(3)$ we find
physical solutions associated to spin-two local excitations.

\end{abstract}.

\maketitle

\section{Introduction}

The discovery of connection variables for general relativity led
to the definition of a new approach for the non-perturbative
quantization of gravity known as loop quantum gravity (LQG)
 \cite{Perez:2004hj,Rovelli:2004tv,Ashtekar:2004eh}. The
introduction of $SU(2)$ connection variables for classical
canonical general relativity  \cite{ash1,barbero}, and the
corresponding use of Wilson loop variables in the quantum theory
 \cite{jac,c8}, allowed the resolution of the longstanding
technical problems that had stopped the development of the {\em
quantum geometro-dynamics} of Dirac, Wheeler, DeWitt among others
 \cite{Rovelli:2000aw}. Among these new achievements are: the
rigorous definition of the kinematical Hilbert space of quantum
gravity, the rigorous quantization of geometric operators such as
area and volume (with the associated prediction of discreteness of
quantum geometry), and the quantization of the highly non linear
Hamiltonian constraint---analog of the Wheeler-DeWitt
equation---governing the dynamics of quantum gravity. The latter
is an important technical achievement of the approach where
background independence and diffeomorphism invariance play a
central role in the elimination of the UV divergences that plague
standard quantum field theories.

Polymer-like excitations known as {\em spin network states} form a
basis of the kinematical Hilbert space $\Hk$. Quantum Einstein's
equations are given by the quantum counterpart of the classical
constraints of canonical general relativity.  A subset of the
constraints---characterized by the vector and Gauss
constraints---requires the physical states of quantum gravity to be
$SU(2)$ gauge invariant and space-diffeomorphism invariant
\footnote{In the Dirac program one starts by defining the so-called
kinematical Hilbert space $\Hk$. One proceeds by representing the set
of classical constraints---here simply denoted by $C\approx 0$---as
quantum operators in $\Hk$. In the classical theory the constraints
generate through the Poisson bracket infinitesimal gauge
transformations; therefore, in the quantum theory $\hat C$ become the
generators of gauge transformations. The Hilbert space of solutions of
the constraint equations $\hat C \Psi=0$ is hence given by the gauge
invariant states and is called the {\em physical Hilbert space},
denoted $\Hp$.}. Since the action of the $SU(2)$ gauge group and
space-diffeomorphism can be unitarily represented in the kinematical
Hilbert space, it is easy to characterize the set of invariant states
and hence the solutions of this subset of quantum constraint equations
by group averaging. Gauge invariant states are given by equivalence
classes of spin network states under diffeomorphisms, i.e., two
polymer-like excitations are regarded as the same if they can de
deformed into each other by the action of a diffeomorphism.

Dynamics is governed by the so-called Hamiltonian constraint,
whose classical form is \be \label{hamm}
H(E_j^a,A_b^i)=\frac{E^a_iE^b_j}{\sqrt{{\rm det}(E)}}F^{k}_{ab}(A)
\epsilon^{ij}_{\ \, k}\approx 0, \ee where $A_a^i$ is an $SU(2)$
connection, $F_{ab}^k(A)$ is its curvature tensor, $E_i^a$ is its
conjugate momentum with the geometric interpretation of a
(densitized) triad field, and we have considered the constraint in
the Riemannian theory (this simplifying assumption will be made
throughout this article). In the quantum theory the Hamiltonian
constraint must be promoted to a quantum operator whose kernel
defines the so-called physical Hilbert space $\Hp$ of quantum
gravity.  The quantization of the Hamiltonian constraint was
introduced by Thiemann in  \cite{th2,th4}. Shortly thereafter it
was pointed out  \cite{marcus} that in addition to (potential)
factor order ambiguities, Thiemann's prescription had an intrinsic
ambiguity labelled by a half-integer $m\in \Z/2$ associated to the
$SU(2)$ unitary representation used to regularize the curvature
tensor $F^{k}_{ab}(A)$ appearing in the classical expression of
the Hamiltonian constraint. In this paper we refer to this problem
as the {\em $m$-ambiguity}.

For every $m\in \Z/2$ one obtains a different quantum Hamiltonian
constraint $\hat H_m$. As argued below, linear combinations of
different regularizations are also good regularizations;
therefore, one obtains an infinite dimensional set of possibly
different theories. In this respect one viewpoint is that the
understanding of the dynamics in each theory would allow the
pinpointing of the correct one by confronting its prediction with
observations. For instance, the analog of the {\em $m$-ambiguity}
appears also in the coupling of quantum gravity with matter. This
ambiguity is known to lead to important physical consequences in
the context of cosmological models inspired by loop quantum
gravity  known as loop quantum cosmology
 \cite{Bojowald:2003uh}. In particular in the evolution of the
universe near the classical big bang
singularity \cite{Bojowald:2003mc}.  These effects are potentially
observable so that comparison with observations is expected to put
constraints on the set of viable theories. Although, this
viewpoint might be argued in the phenomenological framework of
loop quantum cosmology it is not tenable for a fundamental theory
as we will discuss in what follows.

The existence of the $m$ ambiguity is intimately related to the
mechanism leading to the absence of UV problems in loop quantum
gravity. More precisely, in order to regularize quantum operators
corresponding to non linear functionals of the fundamental fields
(e.g. the Hamiltonian constraint) one uses a diffeomorphism
covariant prescription of `point-splitting' consisting of
replacing the connection by holonomies along infinitesimal paths.
The origin of the ambiguity resides in the choice of the $SU(2)$
representation in which these holonomies are taken. Because of
diffeomorphism invariance it turns out that the regulator can be
removed without ever encountering UV divergences. In this way one
ends up with a well defined quantum Hamiltonian constraint, but
only at the price of having an infinite number of consistent but
(in principle) different quantum theories.

The situation is reminiscent of the problem of renormalization in
standard background dependent quantum field theories. There, in
order to make sense of products of operator valued distributions
(representing interactions) one has to provide a regularization
prescription (e.g. an UV cutoff, dimensional regularization, point
splitting, etc.). Removing the regulator is a subtle task
involving the tuning of certain terms in the Lagrangian (counter
terms) that ensure finite results when the regulator is removed.
In fact by taking special care in the mathematical definition of
the `products of distributions at the same point' one can provide
a definition of the quantum theory which is completely free of UV
divergences  \cite{Epstein:1973gw} (see also
 \cite{Scharf:1996zi,Hollands:2001fb,Hollands:2002ux}). However,
any of these regularization procedures is intrinsically ambiguous.
The dimension of the parameter space of ambiguities depends on the
structure of the theory. The right theory must be fixed by
comparing predictions with observations (by the so-called
renormalization conditions). According to this, in loop quantum
gravity one has only achieved the first step: a rigorous
regularization provided by the mathematical framework of the
theory. It remains to settle the crucial issue of how to fix the
associated ambiguities.

According to the previous discussion, ambiguities associated to the UV
regularization allows for the classification of theories in two
important types: {\em renormalizable} and {\em non-renormalizable}
quantum field theories. In a renormalizable theory such as QED there
are finitely many ambiguities which can be fixed by a finite number or
renormalization conditions, i.e., one selects the suitable theory by
appropriate tuning of the ambiguity parameters in order to match
observations. In a non-renormalizable theory (e.g. perturbative
quantum gravity) the situation is similar except for the fact that
there are infinitely many parameters to be fixed by renormalization
conditions. As the latter must be specified by observations, a
non-renormalizable theory has little predictive power.

Removing UV divergences by a regularization procedure is
intimately related to the appearance of ambiguities in the quantum
theory. Although this can happen in different ways in particular
formulations, this problem is intrinsic to the formalism of
quantum field theory. In this respect, it is illustrative to
analyze the non-perturbative treatment of gauge theories in the
context of lattice gauge theory (where the true theory is studied
by means of a regulated theory defined on a space-time
discretization or lattice). It is well known that here too the
regulating procedure leads to ambiguities; the relevance of the
example resides in the fact that these ambiguities resemble in
nature those appearing in loop quantum gravity. More precisely,
let us take for concreteness $SU(2)$ Yang-Mills theory which can
be analyzed non-perturbatively using the standard (lattice) Wilson
action \be S_{\va \rm LYM}=\frac{1}{g^2}\sum_{p}
\left(1-\frac{1}{4}{\rm
Tr}[U_{p}+U_{p}^{\dagger}]\right).\label{two}\ee In the previous
equation $U_{p}\in SU(2)$ is the holonomy around plaquettes $p$,
and the sum is over all plaquettes of a regulating (hyper-cubic)
lattice. It is easy to check that the previous action approximates
the Yang-Mills action when the lattice is shrunk to zero for a
fixed smooth field configuration.  This property is referred to as
the {\em naive continuum limit}. Moreover, the quantum theory
associated to the previous action is free of any UV problem due to
the UV cut-off provided by the underlying lattice.

Is this procedure unique? As it is well known the answer is no.
Among the many ambiguities let us mention the one that, as it will
become clear later, is the closest in spirit to the $m$-ambiguity
in loop quantum gravity. More precisely one can regulate
Yang-Mills theory equally well using the following action instead
of (\ref{two}): \be S^{(m)}_{\va \rm LYM}\propto
\frac{1}{g^2}\sum_{p} \left(1-\frac{1}{2(2m+1)}{\rm Tr}^{\va
(m)}[\Pi^{\va (m)}(U_{p})+\Pi^{\va (m)}
(U_{p}^{\dagger})]\right),\ee where $\Pi^{\va (m)}(U_{p})$ denotes
to the $SU(2)$ unitary irreducible representation matrix (of spin
$m$) evaluated on the plaquette holonomy $U_{p}$. Or more
generally one can consider suitable linear combinations \be S_{\va
\rm LYM}=\sum_{m}\ a_m \ S^{(m)}_{\va \rm LYM}. \ee From the view
point of the classical continuum theory all these actions are
equally good as they all satisfy the {\em naive continuum limit}.
Do these theories approximate in a suitable sense the continuum
quantum field theory as well? and are these ambiguities
un-important in describing the physics of quantum Yang-Mills
theory? The answer to both of these questions is yes and the
crucial property that leads to this is the renormalizability of
Yang-Mills theory. Different choices of actions lead indeed to
different discrete theories.  However, in the low energy effective
action the differences appear only in local operators of dimension
five or higher. A simple dimensional argument shows that in the
continuum limit (i.e. when the regulating lattice dependence is
removed by shrinking it to zero) all the above theories lead to
the same predictions in the sense that can safely ignore
non-renormalizable contributions. Therefore, the ambiguities at
the {`microscopic level'} do not have any effect at low energies
where we recover quantum Yang-Mills theory.

The situation in LQG looks at first quite similar. In order to
quantize the Hamiltonian constraint one also needs to make
mathematical sense of the highly non linear (not even polynomial)
form of the Hamiltonian constraint (\ref{hamm}). The Hamiltonian
constraint is quantized by means of a regularization procedure
that, due to the manifest background independence of the approach,
does not lead to any UV divergencies when removed (no hidden
infinities are ever encountered). However, as in standard QFT
ambiguities arise as a consequence of the regularization. Here we
are concerned with what we have called the $m$-ambiguity which
appears when non linear functions of the connection $A$ are
replaced by holonomies in the regularization of the Hamiltonian
constraint (\ref{hamm}). The $m$-ambiguity is associated (in
analogy to the previous example in the context of lattice gauge
theory) with the $SU(2)$ representation chosen in the
regularization. As a consequence one obtains an $m$-worth ($m\in
\Z/2$) of (smeared)\footnote{The smeared Hamiltonian constraint is
defined as
\[H[N]=\int \frac{N(x)E^a_i(x)E^b_j(x)} {\sqrt{{\rm
det}(E(x))}}F^{k}_{ab}(A(x)) \epsilon^{ij}_{\ \, k}\ d^3x, \] and
$N(x)$ is a scalar test function called the lapse.} quantum
Hamiltonians, $\hat H_m[N]$, that are consistent in the sense of
Thiemann. More generally any linear combination
\be\label{haha}\hat H[N]=\sum_m\ a_m \ \hat H_m[N] \ \ \ {\rm
with} \ \ \ \sum_m a_m=1\ee is also a consistent quantization. The
nature of this ambiguity is very similar to the example considered
in the context of lattice gauge theory above but the naive
implications seem rather dangerous in the case of gravity.

If one would argue in analogy to the lattice gauge theory case one
immediately runs into trouble because of the non-renormalizability
of gravity. Indeed for gravity the non trivial information about
the quantum theory is encoded in the dimension five and higher
local operators in the effective action (i.e. the infamous
higher curvature quantum corrections to the Einstein-Hilbert
action). Consequently, and according to our previous argument,
these are precisely the terms that would be affected by the
ambiguities of the microscopic theory, and one would need to
perform an infinite set of independent measurements in order to
fix the ambiguities of the fundamental theory. Such a scenario
would place the non perturbative approach of LQG at the same
footing as the standard perturbative approach in the sense of
predictive power.

However, one should doubt of the validity of the previous argument
on the basis that it is constructed from a notion of `continuum limit'
which is only applicable to background dependent theories. For
example in lattice gauge theories it is relatively easy to define
the notion of a continuum limit by simply studying the dependence
of the observables of the theory as a function of the lattice
constant. Due to background independence there is no analog of the
lattice constant in loop quantum gravity. Geometry is dynamical
and the only scale entering the theory is the fixed Planck length
that modulates the spectrum of geometric operators. Due to both
technical and conceptual difficulties associated to the definition
of the continuum in  LQG, an explicit treatment of the question of
the effects of the ambiguities at low energies is not possible at
this stage. There are indeed indications that the low energy limit
in a background independent theory is very different from what one
would naively hope from the experience in standard QFT
 \cite{daniel}. However, even though it may be wrong to use the
heuristics of standard QFT, this perspective poses a genuine
question that requires an answer. The goal of this paper is to
shed some light onto this important issue.

It is interesting to notice that in the simplified context of loop
quantum cosmology one can study the effects of the $m$-ambiguity,
and arrive to conclusions that are in agreement with the previous
motivation. Even though, in this framework, one deals with
finitely many degrees of freedom, the ambiguities of the full
theory are inherited by the model due to the particular way in
which the model is derived from the full theory. In this
simplified setting one can compute quantum corrections to the
classical theory in the sense of an effective theory. These appear
in fact as higher curvature corrections to Hamiltonian constraint
 \cite{Willis:2004br}. The precise form of these corrections
depends indeed on the value of the parameter $m$
 \cite{Vandersloot:2005kh}. As in the previous case one should
interpret these results with due care. In particular loop quantum
cosmology is not a fundamental description of quantum gravity, and
it is not even diffeomorphism invariant. Nevertheless, it provides
an new perspective to arrive at the key question that motivates
this work.

Finally, it is also possible that some set of the ambiguities
found in the quantization of the Hamiltonian constraint are of no
physical relevance due to consistency conditions that can already
be found by studying in more detail the dynamics of the theory. If
that is the case then there is a chance that we can shed some
light on the issue before completely resolving the problem of the
low energy limit of LQG. This is in fact the avenue that will be
explored in this work.

These considerations confront loop quantum gravity with two
obvious alternatives:
\\

\noindent {\bf i)} From the infinite dimensional set of quantum
Hamiltonians (\ref{haha}) only a finite dimensional subset leads
to mathematically consistent and physically different theories.
\\

\noindent {\bf ii)}  The infinite dimensional set of quantum
Hamiltonians (\ref{haha}) leads to an infinite dimensional space
of mathematically consistent and physically different theories.
\\

The possibility i) is desirable while possibility ii) is
equivalent to the status of perturbative quantum gravity in the
sense of predictive power. Despite its central role in
understanding the theory of LQG, this question has been only
marginally posed  \cite{Nicolai:2005mc}. We will explicitly show
that in the case of 2+1 gravity the first possibility holds.  In
fact there are an infinite dimensional set of mathematically
consistent regularizations of the constraints of 2+1 gravity, yet
they all lead to the same physical theory. In this case we are
free to choose the simplest one which corresponds to using the
fundamental representation in the regularization of the curvature
constraint.

The fact that 2+1 gravity is a topological quantum field theory might
indicate that there cannot be any UV related renormalization problems
because the theory has no local degrees of freedom.  However, we must
emphasize that before implementing the curvature constraint the
kinematical Hilbert space of the theory corresponds to the kinematical
Hilbert space of LQG, and thus that of a full fleshed field
theory. Hence, ambiguities arise in the definition of the dynamics in
a way that mimics the four dimensional case.  If these ambiguities
should also desappear in 3+1 gravity we might expect to learn
something about the underlying mechanism by studying how it happens in
2+1 dimensions.

Indeed, using the insight from the 2+1 theory we provide evidence to support a
similar conclusion in 3+1 quantum gravity. Our result in 3+1 gravity
is weaker due to the present lack of suitably defined notion of
physical inner product. Our analysis will be performed in the
Riemannian theory in the framework of Thiemann's quantization;
however, the general ideas presented here are expected to be relevant
for other prescriptions for the definition of the quantum
dynamics---such as the {\em master constraint program}
 \cite{Thiemann:2003zv} or that of {\em consistent discretizations}
 \cite{Gambini:2004vz,Gambini:2005za,Gambini:2005pg,Gambini:2005sv}---where
the same regularization ambiguity arises. This work is meant to
provide a direction that could lead to a possible resolution of the
ambiguity issue. A stronger result (as the one in 2+1 gravity) would
require explicit knowledge of the (yet not available) notion of
physical probability derived from the theory.

Are there other ambiguities?  Perhaps the most obvious ambiguity
in the quantization of the classical expression (\ref{hamm})
concerns the ordering of the densitized triad fields and the
connection. However, background independence and consistency with
the (recently shown to be unique  \cite{lost}) kinematical
structure of loop quantum gravity appears to drastically reduce
factor ordering ambiguities. The only known mechanism for the
quantization of the non linear $E$-dependent part of the
Hamiltonian constraint is due to Thiemann and based on the
observation that one can write (\ref{hamm}) as \be H \propto
\epsilon^{abc}\epsilon_{ijk}e_a^iF_{bc}^{jk}\ \ \ \mbox{with} \ \
\ e_a^i(x)= \{A_a^i(x),\int dy^3 \ \sqrt{|{\rm det}(E)|}\}.  \ee
The previous expression is used in the quantization where the
Poisson bracket is promoted to a commutator and the integral
corresponds to the well understood quantum volume operator (there
are in fact two proposed versions of the latter  \cite{ash22,c3};
however, resent results indicate that only one appears to be
consistent  \cite{Giesel:2005bm,Giesel:2005bk}). Because Thiemann's
prescription requires  the $E$ dependent part of the Hamiltonian
to be treated in this way we end up with only two factor ordering
possibilities: $\hat e_a^i$ on the left or on the right of $\hat
F(A)$. The first of the previous possibilities does not lead to a
well defined operator  \cite{thiemann}, technically it is ruled out
by cylindrical consistency  \cite{Thiemann:1996aw}. So we conclude
that factor ordering seem not to be a source of infinitely many
ambiguities.  Another ambiguity noticed in the literature is
associated to, what we can call, the `combinatorial' possibilities
in the regularization of the curvature part of (\ref{hamm}). As we
mentioned above one regularizes the connection dependence in the
Hamiltonian by using holonomies. There at least two natural
choices: one where the action of the Hamiltonian constraint on a
spin network creates new nodes, and the other where only the
valence of nodes is altered by the action of the constraint but no
new nodes are created.  A  new manifold of ambiguities appears if
one considers the coupling of gravity to
matter \cite{Bojowald:2004xq,Bojowald:2002ny}. The effects of these ambiguities
will not be studied here. We concentrate on the $m$-ambiguity
which gives rise to infinitely many a priory consistent theories
and is most clearly related to the regularization procedure.

\section{The m-ambiguity in quantum canonical 2+1 Riemannian gravity}

A complete account of the canonical quantization of 2+1 gravity using
LQG techniques is provided in  \cite{Noui:2004ja}; we will follow the
notation therein. If one starts from the kinematical Hilbert space $\Hk$
spanned by spin network states the only remaining constraint in 2+1
gravity is the quantum curvature constraint
\[\hat F(A)|\psi>=0.\]
The physical inner product and the physical Hilbert space, $\Hp$,  of 2+1
gravity can be defined by introducing a regularization of the formal
expression defining the generalized projection operator into the
kernel of $F$, namely,
\begin{equation}
P =``\prod_{x \in \ \Sigma} \delta(\hat F(A))"=\int D[N] \ {\rm
exp}(i\int \limits_{\Sigma} {\rm Tr}[ N \hat {F}(A)])
\;,\label{ppp}
\end{equation}
where $N\in su(2)$, and $\Sigma$ denotes the $2$-dimensional
Riemann surface representing space. In  \cite{Noui:2004ja} it is
shown how the previous object can be given a precise definition
leading to a rigorous expression for the physical inner product of
the theory. However, in order to give a precise meaning to the
previous formal expression it is necessary to introduce a
regularization as an intermediate step for the quantization due to
the non-linear dependence of the constraint on the fundamental
variables. In this section we observe that the analog of the {\em
$m$-ambiguity} in 3+1 gravity appears when the regulator is
introduced. Therefore we first generalize the construction of
 \cite{Noui:2004ja} to this case.

In order to motivate the regularization consider a local patch
$U\subset \Sigma$ where we choose the cellular decomposition to be
square with cells of coordinate length $\epsilon$. In that patch, the
integral in the exponential in (\ref{ppp}) can be written as a Riemann
sum \be \label{FF} F[N]=\int\limits_{U} {\rm Tr}[ N {F}(A)]=
\lim_{\epsilon\rightarrow 0}\ \sum_{p} \epsilon^2 {\rm Tr}[N_{p}
F_{p}],\ee where $p$ labels plaquettes, $N_{p}\in su(2)$, and
$F_{p}\in su(2) $ are values of $N^i\tau_i$, $\tau_i
\epsilon^{ab}F^{i}_{ab}[A]$ at some interior point of the plaquette
$p$, and $\tau_i$ are the generators of $su(2)$. The tensor
$\epsilon^{ab}$ is the $2$-dimensional Levi-Civita tensor. The
quantity $F[N]$ corresponds to the smeared curvature constraint.

The basic observation is that given the holonomy $U_{p}\in SU(2)$
around the plaquette $p$ and a unitary irreducible
representation of $SU(2)$,  $\Pi^{\va (m)}$,  one can write
\[\Pi^{\va (m)}[U_{p}]={1}^{\va (m)} + \epsilon^2 F^i_{p}(A) \tau^{\va (m)}_i+{\cal O}(\epsilon^2),\]
where ${1}^{\va (m)}$ is the identity in the
representation $m$ and $\tau^{\va (m)}_i$ is the $i$-th generator
in the corresponding representation.  Which implies \be \label{ff}
F[N]=\int\limits_{U} {\rm Tr}[ N {F}(A)]=\lim_{\epsilon\rightarrow
0}\ \sum_{p} \frac{{\rm Tr}^{\va (m)} [N_{p} \Pi^{\va
(m)}[U_{p}]]}{C^{\va (m)}}\;,\ee where the ${\rm Tr}^{\va (m)}$ in
the r.h.s. is taken in the representation $m$, $N_{p}=N^k_{p}
\tau^{\va (m)}_k $ and $C^{\va (m)}={\rm Tr}^{\va (m)}[\tau^{\va
(m)}_3\tau^{\va (m)}_3]$. Notice that the explicit dependence on
the regulator $\epsilon$ has dropped out of the sum on the r.h.s.,
a sign that we should be able to remove the regulator upon
quantization. The r.h.s. can be easily promoted to a sum of self
adjoint operators acting in the kinematical Hilbert space, so the
previous prescription provides a half-integer-worth of
quantizations of $F[N]$ in the sense of Rovelli-Gaul  \cite{marcus}
(the operator $\hat \Pi^{\va (m)}[U_{p}]$ acts simply by
multiplication in $\Hk$  \cite{lost}). The use of holonomies in the
quantization of $F[N]$ (which is the natural point-split-like
regularization adapted to the kinematical structure of the theory)
is responsible for the occurrence of the $m$-ambiguity.

Following  \cite{Noui:2004ja} one introduces ${\rm P}^{\va
(m)}_{\epsilon}$---a regularization of the generalized projection
operator in terms of the representation $m$---in terms of a definition of its matrix elements between elements of the spin network basis denoted $\{|s>\}$, namely\ba
\label{twenty} \ssp & = & \lim_{\epsilon\rightarrow 0} \ \
<\prod_{p} \ \int \ dN_{p} \ {\rm exp}(i  \frac{{\rm Tr}^{\va (m)}
[N_{p}\hat
\Pi^{\va (m)}[U_{p}]]}{C^{\va (m)}}) s, \; s^{\prime}>\\
\nonumber & = & \lim_{\epsilon\rightarrow 0} \ \ <\prod_{p} \
{d^{\va (m)}}(U_{p})s, \; s^{\prime}>,\label{P3} \ea where in the
last equation we have introduced the distribution $d^{\va(m)}(U)$
that we formally write as \be \label{dd} d^{\va(m)}(U)=\int dN \
{\rm exp}(i  \frac{{\rm Tr}^{\va (m)} [N \hat \Pi^{\va
(m)}[U]]}{C^{\va (m)}}) = C^{{\va (m)}3}\delta({\rm Tr}^{\va (m)}
[\tau^{\va (m)}_1 \Pi^{\va (m)}[U]]) \delta({\rm Tr}^{\va (m)}
[\tau^{\va (m)}_2 \Pi^{\va (m)}[U]]) \delta({\rm Tr}^{\va (m)}
[\tau^{\va (m)}_3 \Pi^{\va (m)}[U]]). \ee It is easy to check that
$d^{\va (1/2)}(U)=\delta(U)$, i.e., the delta distribution on
$SO(3)$ directly from the $N$ integration. Therefore, $d^{\va
(1/2)}(U)$ projects into the identity in the sense that $\int dU
f(U)d^{\va (1/2)}(U)=f({1})$. However, for $m\not=1/2$ the
group averaging is more subtle and the r.h.s. of the previous
equation is not well defined as a distribution. We will give a
precise definition of $d^{\va(m)}(U)$ for $m\not=1/2$ below. The
properties of $d^{\va(m)}(U)$---as shown in
 \cite{Noui:2004ja}---completely determine the physical scalar
product of the theory. In fact the above property of $d^{\va
(1/2)}(U)$ implies that ${\rm P}^{\va (1/2)}$  defines a
projection operator into flat-connection-configurations and
therefore yields a physical Hilbert space corresponding to
finitely many topological degrees of freedom.

Before studying the case $m\ge 1$ we will illustrate the main idea
in a simpler case: three dimensional BF theory with internal gauge
group $G=U(1)$. This example illustrates the main idea that we
will be applied in the rest of the paper. The analog of equation
(\ref{dd}) is given by the expression\footnote{The analogy is self
evident observing that ${\rm Tr}^{\va (m)} [N_{p}\hat
\Pi^{\va (m)}[U_{p}]]={\rm Tr}^{\va (m)} [N_{p}(\hat \Pi^{\va
(m)}[U_{p}]-{1}^{\va (m)})]$ from the fact that $N_{p}$ is
traceless.} \be \label{trtr} d^{\va(m)}(\phi)=\int dN \ {\rm
exp}(iN[e^{i m \phi}-1])=\delta[e^{i m \phi}-1], \ee which we can
expand in terms of $U(1)$ unitary irreducible representations as
\be d^{\va(m)}(\phi)=\sum_k c_k^{\va (m)} \ e^{i k \phi},  \ee
where \be c_k^{\va (m)}=\int \frac{d\phi}{2\pi} \
e^{-ik\phi}\delta[e^{i m
\phi}-1]=\frac{1}{2\pi|m|}\sum_{\alpha=1}^{m-1}
e^{-ik\phi_{\alpha}}, \ee where $\phi_{\alpha} = \alpha \delta$
with $\delta=2\pi/m$ are the roots of the argument of the delta
function above. These roots are the solutions to the regularized
constraint $F={\rm exp}(i m\phi)-1$. We see that as a consequence
of our regularization the constraint admits extra solutions in
addition to the flat one $\phi=0$. The sum corresponds to a
geometric sum, namely \be c_k^{\va
(m)}=\frac{1}{2\pi|m|}\sum_{\alpha=1}^{m-1} \left[e^{-i k
\delta}\right]^{\alpha}=\frac{1-e^{-i k \delta m}}{1-e^{-i k
\delta m}}= \left\{\begin{array}{ccc}(2\pi)^{-1} \ \ \ \forall \ \
k=p \ m, \ \ \ p\in\Z\\ \!\!\!\!\!\!0 \ \ \ \ \ \ \ \ \ \ \ \ \ \
\ \ \ {\rm otherwise}
\end{array}\right.
\ee If we proceed as in   \cite{Noui:2004ja} we would find that
unless $m=1$ (i.e. the fundamental representation) we would obtain
a theory with infinitely many degrees of freedom. This is because
the vanishing of infinitely many Fourier components of $d^{\va
(m)}(\phi)$ for $m\not=1$ implies a reduction of the space of
zero-norm states with respect to the $m=1$ quantization. Hence,
the physical Hilbert space becomes larger. The argument presented
here is rather formal. This is because the $U(1)$ case presents
some extra subtleties at the time of defining the physical inner
product which are not present in the non-Abelian case which will
be treated in more detail in the following section \footnote{When
the fundamental representation is used in the regularization, the
vacuum to vacuum physical transition amplitude if the theory is
defined on $M=\Sigma\times \R$ with $\Sigma$ given by a Riemann
surface of genus $g$ is given by $<P,1>=\sum_k
\Delta_k^{2-2g}$---where $\Delta_k$ is the dimension of the
irreducible unitary representation $k$---which is convergent for
the $SU(2)$ case and $g>1$ but always ill defined for $U(1)$. This
is an example of the kind of technical difficulties we would
encounter if we would like to completely analyze the $U(1)$
case.}. The choice of $m>1$ introduces {\em spurious} solutions to
the regularized constraints.

Now we essentially repeat the previous derivation for $SU(2)$, but
we go further removing the regulator and  constructing in this way the
physical inner product. We shall see that the spurious solutions
appearing in the previous example are also present in 2+1 gravity
for certain bad regularizations. We will show that for these
choices the regulator cannot indeed be removed and such
regularizations must be ruled out as inconsistent. This will lead
to a unique theory in the case of three dimensional gravity.

Let us analyze $d^{\va(m)}(U)$ defined in equation (\ref{dd}) in
more detail. The simplest way is to use the isomorphism between
$SU(2)$ and $S^3$. Any element $U\in SU(2)$ can be written as
\be\label{12} U=x^{\mu}\tau_{\mu} \ \ \ {\rm where} \ \ \
x^{\mu}x^{\nu}\delta_{\mu\nu}=1, \ee $\mu,\nu=1,\cdots,4$, and
$\tau_{0}=1$ and $\tau_{\alpha}=i\sigma_{\alpha}$ for
$\sigma_{\alpha}$ the Pauli matrices for $\alpha=1,2,3$. In terms
of this parametrization of $SU(2)$ one can write the unitary
irreducible representations of spin $m$ as \be \Pi^{\va
(m)}[U]^{A_1\cdots A_{2m}}_{B_1\cdots B_{2m}}=\ x^{\mu_1}\cdots
x^{\mu_{2m}}\ \tau^{(A_1}_{\mu_1\ B_1} \cdots \
\tau^{A_{2m})}_{\mu_{2m}\ B_{2m}}, \ee from where it follows that
\be \label{ee}{\rm Tr}^{\va (m)} [\tau^{\va (m)}_i \Pi^{\va
(m)}[U]]=x^i Q^{(2m-1)}(x^{\mu}), \ee where $Q^{(2m-1)}(x^{\mu})$
is a polynomial of degree $2m-1$. The fact that
$d^{\va(m)}(U)=d^{\va(m)}(gUg^{-1})$ implies that
$Q^{(2m-1)}(x^{\mu})$ is rotational invariant as a function of
$\vec x$ or, equivalently, only dependent on $x^0$ (using
(\ref{12})). Therefore, only for $m=1/2$ the only solution to the
three traces in (\ref{P3}) equal to zero is \ $\vec x=0$ which
implies $U=1$. However, for $m\ge 1$ in addition to $\vec x=0$ we
have the roots of $Q^{(2m-1)}(x^{\mu})=0$. For example for $m=1$
one has \be Q^{(1)}(x^{\mu})=2 x^0=2\sqrt{1-\vec x \cdot \vec x}.
\ee In this case equation (\ref{ee}) vanishes for the point $\vec
x=0$ and the 2-sphere $\vec x \cdot \vec x=1$. The fact that the
configurations that solve the constraint is the union of
sub-manifolds of $SU(2)$ with different dimensions (the point
$x^0=1$ and the sphere $|\vec x|=1$) implies that (\ref{dd}) is
ill-defined  as a distribution. In order to carry on one has to
introduce a regularization having in mind that $d^{\va (m)}[U]$
must project onto the identity $x^0=1$ and the each of the
2-spheres of $S^3$ that are solutions of $Q^{(2m-1)}(x^{\mu})=0$.
The regularization procedure is ambiguous. The ambiguity can be
parameterized by two parameters, namely \be\label{16}
d^{\va(m)}(U[x^{\mu}])=\lambda_1
\prod_{i=1}^{3}\delta(x^{i})+\lambda_{2}
\delta(Q^{(2m-1)}(x^{\mu})). \ee Notice that if we would choose
$\lambda_2=0$ we would immediately reproduce the standard
quantization based on the fundamental representation. Since our
aim is to explore the possibility of constructing a theory which
is both well defined but different from the one obtained for
$m=1/2$ we proceed by assuming that $\lambda_2\not=0$.

As in reference   \cite{Noui:2004ja} it will be convenient to
expand the distribution $d^{\va(m)}(U)$ in terms of unitary
irreducible representations. This allows us to write the plaquette
contributions (\ref{twenty}) in terms of sums over Wilson loops
which can be easily represented by self adjoint operators in
$\Hk$. More precisely we want \be d^{\va(m)}(U)=\sum_j c^{\va
(m)}_j \ {\chi}_j[U], \ee where ${\chi}_j(U)$ is the character or
trace of the $j$-representation matrix of $U\in SU(2)$ and the
coefficients $c_j^{\va (m)}$ are given by the Peter-Weyl theorem, namely \be \label{18} c_j^{\va
(m)}=\int dU\ d^{\va(m)}(U) {\chi}_j[U^{-1}]= \frac{1}{\pi^2} \int
dx^{\mu} \delta(x^{\mu}x^{\nu}\delta_{\mu\nu}-1) \ d^{\va(m)}\
(U[x^{\mu}]) {\chi}_j[U^{-1}[x^{\mu}]] ,\ee where the integration
is performed with the Haar measure of $SU(2)$ that, in the
coordinates we are using, takes the simple form
\[d\mu_{\va H}=\pi^{-2} dx^{\mu} \
\delta(x^{\mu}x^{\nu}\delta_{\mu\nu}-1).\] For $m=1/2$ we obtain the
familiar result $c_j^{\va (1/2)}=2j+1$ for $j\in \Z$
 and zero otherwise, i.e. $d^{1/2}[U]$ is the $SO(3)$
  delta distribution \footnote{In fact one cannot get
the mode expansion of the $SU(2)$ delta function coming from the
integral definition of $d^{1/2}[U]$. In order to have the
half-integer representations in 2+1 gravity one must define
$c_j^{\va (1/2)}=2j+1$ for all half-integers   \cite{thesis}.}. For
$m=1$, $Q^{(1)}(x^{\mu})=2 x_0$; therefore using (\ref{16}) and
(\ref{18}) we obtain \be c_j^{\va (1)}=\lambda_1 (2j+1)+
{\lambda_2} \chi_j[U_0], \ee where $U_0$ is in the conjugacy class
of the element labelled by coordinates  $x^0=0,x^1=x^2=0,x^3=1$.
Since in the expression of the physical inner product we can
absorb an overall factor, we will define \be c_j^{\va (1)}=
(2j+1)+ \lambda \ \chi_j[U_0]=(2j+1)\left[1+\lambda
\frac{\sin[(2j+1)\frac{\pi}{2}]}{(2j+1)}\right], \ee where
$\lambda$ parameterizes the remaining ambiguity.

The previous equation allow us to write the distribution
$d^{(1)}[U]$ as a sum of holonomy operators in the corresponding
irreducible representations. We can represent the regulated
projector as a sum of product of such fundamental Wilson loops
based on the plaquettes of the regulating lattice. In order to
complete the definition of the theory we must take the limit
$\epsilon\rightarrow 0$ in the definition of the physical inner
product. This amount to shrinking to zero the cellular
decomposition of $\Sigma$ used as regulator of $\rm P$. To make
our point it will be sufficient to consider the vacuum-to-vacuum
transition amplitude defined by equation (\ref{twenty}) when the
states $|s>=|s^{\prime}>=|1>\in \Hk$.

\begin{figure}[h!!!!!]
 \centerline{\hspace{0.5cm} \( \sum \limits_{j k} c^{\va (m)}_j  c^{\va (m)}_k
\begin{array}{c}
\includegraphics[height=3cm]{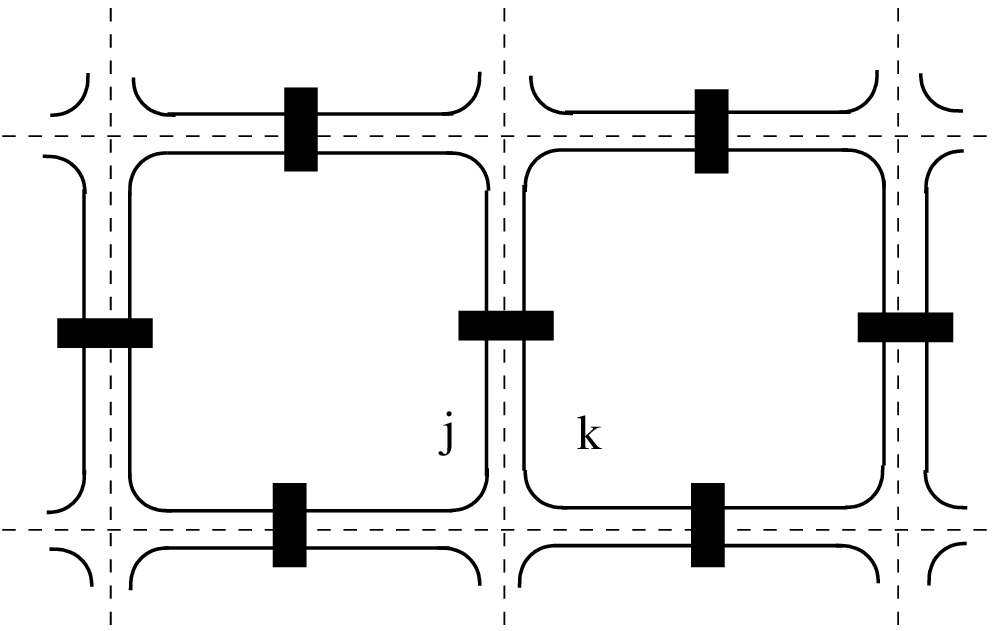}
\end{array} =  \sum \limits_k  \frac{[c^{\va (m)}_k]^2}{2k+1}
\begin{array}{c}
\includegraphics[height=3cm]{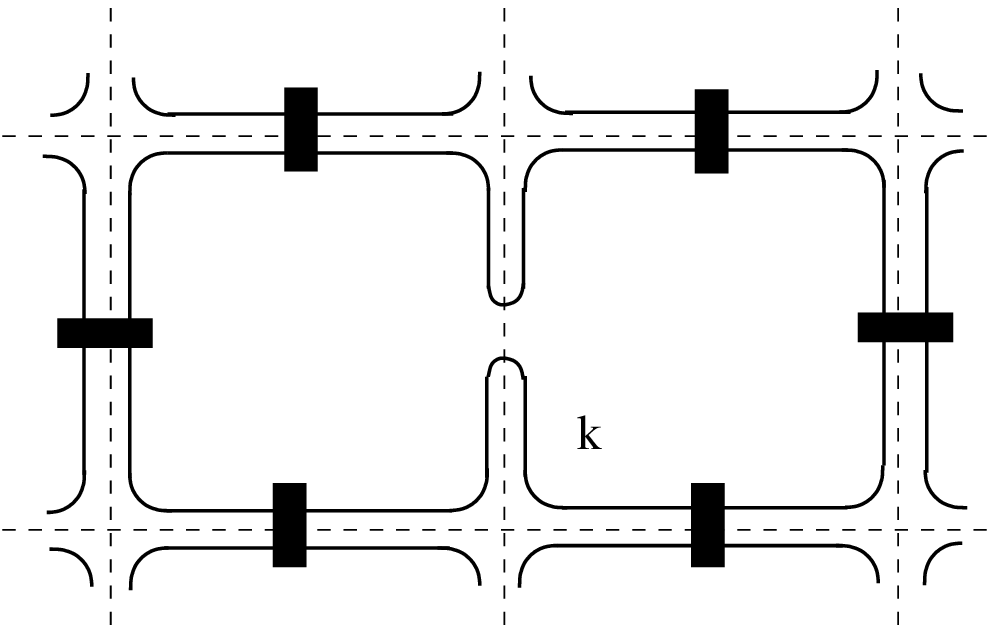}
\end{array}
\) }
\caption{\small Infinitesimal plaquette-delta-distributions can be
  integrated and fussioned with the corresponding modification of the
  face amplitude. The lines in the previous figure represent the
  holonomy around plaquettes of the regulator in the representation
  denoted by the latin index $k$ and $j$ in this case.  the dark boxes
  denote integration of the generalized connection associated to the
  corresponding edge.  The previous equation is a trivial consequence
  of the orthogonality of unitary irreducible representations of
  $SU(2)$. The plaquettes in the picture are square as a matter of
  simplicity.}
\label{hh}
\end{figure}

In the case $m=1/2$ the limit $\epsilon\rightarrow 0$ is
straightforward because the integration of the connection on the
boundary of neighboring plaquettes is, in that case, simply equivalent
to a fusion of plaquettes with no change in the amplitude (see
Figure~\ref{hh} with $c_j^{\va(1/2)}=2j+1$).  In this sense we have a
trivial scaling or renormalization of the amplitudes for $m=1/2$ so
that the continuum limit produces a topological quantum field theory
(for details see  \cite{Noui:2004ja}). In fact the vacuum-to-vacuum
amplitude is \be
\label{topo}<{\rm P}^{\va (\frac{1}{2})},1>=\sum_j (2j+1)^{2-2g},\ee where $g$ is the genus of
$\Sigma$.

For an arbitrary $m$ the situation changes radically.  This can be
illustrated in our present example $m=1$. Because of the additional
solution $U_0\not=1$, the fusion move does no longer imply a trivial
renormalization of the face amplitude (see Figure \ref{hh}).
Integrating over all the internal connections we obtain
\be <{\rm P}^{\va (1)}_{\epsilon},1>=\sum_j
(2j+1)^{2-2g} \ \left[1+\lambda
\frac{\sin[(2j+1)\frac{\pi}{2}]}{(2j+1)}\right]^{\frac{A}{\epsilon^2}},
\ee where $<{\rm P}^{\va (1)}_{\epsilon},1>$ denotes the vacuum-to-vacuum amplitude
before the regulator has been removed, $A$ is the coordinate area of $\Sigma$, and $A/\epsilon$ is the number
of plaquettes in the cellular decomposition.

For $\lambda\not=0$ the limit $\epsilon\rightarrow 0$ of the
previous face amplitude is ill defined. For constant $\lambda$ the
face amplitude will either diverge or converge to zero depending
on the value of the representation $j$. In order to avoid this
problem we could renormalize $\lambda$ as we shrink the lattice.
For instance the limit would in fact be well define we chose
$\lambda=\epsilon^{-2}\lambda_0$. In this case we get \be <{\rm
P}^{\va (1)},1>=\lim_{\epsilon \rightarrow 0}<{\rm P}^{\va
(1)}_{\epsilon},1>=\sum_j (2j+1)^{2-2g}\ {\rm exp}\left[\lambda_0
A \frac{\sin[(2j+1)\frac{\pi}{2}]}{(2j+1)}\right]. \ee The
previous amplitude explicitly depends on the coordinate area of
$\Sigma$. We can insist in defining the limit by renormalizing the
ambiguity parameter but at the cost of loosing background
independence. It is clear the the theory obtained for $m=1$ has
nothing to do with 2+1 quantum gravity. In other words we have
taken the limit $\epsilon\rightarrow 0$ but the result is not even
diffeomorphism invariant: it remains in the dependence of the
amplitude on the coordinate area. We have run into an anomaly of
the kind described in  \cite{myo}. We can avoid the previous
problem if we chose $\lambda(\epsilon)={\cal O}(\epsilon^2)$. In
that case we would recover the topological amplitude (\ref{topo})
as in the case $m=1/2$. This is not surprising as we are simply
suppressing any contribution of the spurious solutions in the
continuum limit.

For $m=3/2$ the situation is simply the same, which illustrates the
generic case. In this case $Q^{(2)}(x^{\mu})=8 x^ix_i-20 (x^0)^2$. A
similar analysis gives therefore we get \be c_j^{\va (3/2)}=\lambda_1
(2j+1)+ {\lambda_2} \chi_j[U_0] \ee where $U_0$ is in the conjugacy
class of the element labelled by the point
$x^0=\sqrt{2/7},x^1=x^2=0,x^3=\sqrt{5/7}$.  As before (and for any
$m\ge 1$) the presence of spurious solutions would spoil the existence
of a diffeomorphism invariant continuum limit unless the physical
inner product is defined in such that the extra solutions have zero
physical norm. In that case the theory obtained coincides with the one
constructed in terms of the fundamental representation $m=1/2$.

\subsection{Linear combinations}\label{lc}

The problem with the quantization of the curvature constraint in
terms of a single representation $m$ that is different from the
fundamental one can be traced back to the existence of non trivial
configurations that solve the regulated constraint. These extra
solutions do not correspond, in classical terms, to $F=0$. In the
limit $\epsilon \rightarrow 0$ the spurious solutions define wild
oscillatory configurations at the coordinate scale set by
$\epsilon$. These solutions conspire to make the elimination of
the regulator ill defined. We have seen in the previous section
that unless the spurious solutions are appropriately suppressed
(which leads to the quantum theory obtained for $m=1/2$) the
continuum limit does not exist or is anomalous.

One can avoid the previous undesired effect by considering those
{\em good} regularizations that do no introduce spurious
solutions. In fact this can be easily characterized as follows:
Instead of using a regularization consisting on a single
irreducible representation one can study the general case where
the curvature constraint is quantized by an arbitrary linear
combination of Wilson lines in any representation.  Namely we
replace (\ref{ff}) by \be
 \hat F[N]=\sum_{m} a_m \hat F^{\va(m)}[N]=\lim_{\epsilon\rightarrow 0}\ C^{-1} \sum_{p} \ \sum_{m } a_m\  {\rm Tr}^{\va
(m)} [N_{p}\hat \Pi^{\va (m)}[U_{p}]]\;, \ee where $C^{-1}$ is the
appropriate normalization factor for $\sum_m a_m=1$.

There exists an infinite dimensional space of such regularizations,
parameterized by the coefficients $\{a_m\}$. From this infinite
dimensional set of theories only those which satisfy \be \sum_{m }
a_m\ {\rm Tr}^{\va (m)} [N_{p} \Pi^{\va (m)}[U]=0 \ \ \ {\rm
iiff}\ \ \ U={1} \ee lead to theories where the continuum
limit is well defined. In fact the individual values of the
coefficients $\{a_m\}$ plays no physical role, and as long as the
previous equation holds the corresponding physical inner product
is unique.

In the $U(1)$ example this corresponds to any periodic function
$F[\phi]$ on the interval $[0,2\pi]$ vanishing at $0$. It is
obvious that there is an infinite dimensional space of such
functions. The analog of Equation (\ref{trtr}) becomes \be
d(\phi)=\int dN \ {\rm exp}(iNF[\phi])=F^{\prime}[0] \delta[\phi].
\ee Except for a trivial overall factor {\em renormalization} we
obtain the result that follows from the quantization based on the
fundamental representation. The result is exactly the same in the
non-Abelian case. So we conclude that considering arbitrary linear
combinations of representations we can obtain well defined
quantizations of 2+1 gravity. However, the resulting theory is
completely equivalent to the $m=1/2$ quantization. We are in fact
in the situation (i) described in the introduction.

\subsection{Covariant spin foams}\label{sfm}

At this stage it should be clear that the analysis presented above
can be extended with mild modifications to the covariant picture.
More precisely in the lattice definitions of the path integral for
2+1 quantum gravity that leads to the Ponzano-Regge model one can
also study the effect of the modification of the simplicial action
by replacing the customary regularization of the curvature tensor
in terms of the Wilson line in the fundamental representation by
an arbitrary function of the holonomy around plaquettes satisfying
the naive continuum limit property.

In the case of a regularization based on a single unitary
representation; for $m\not=1/2$ discretization independence of the
partition function is lost and the path integral is no longer well
defined. The continuum limit is lost. The good regularizations are
characterized as in the previous section and are equivalent to
that defined in terms of the fundamental representation, in terms
of which we recover a unique result: the standard Ponzano-Regge
model. Notice that this can also be interpreted from the point of
view developed in \cite{myo}, if the spin foam face amplitude is
not equal to the dimension of the representation labelling the face,
the spin foam amplitudes are not well defined in the equivalence
classes of spin foams and hence are regarded as anomalous.

\section{The {\em $m$-ambiguity} in 3+1 gravity}

In 3+1 gravity our strategy is similar to that of 2+1 gravity. We will
show that unless $m=1/2$ (for $SU(2)$)---or $m=1$ (for $SO(3)$)---is
used in the regularization of the Hamiltonian constraint, the
resulting theory contains spurious local degrees of freedom. These are
the analog of the new solutions found above which interfere with the
existence of the continuum limit in 2+1 gravity. We will explicitly
demonstrate the existence of such solutions in 3+1 gravity by
constructing explicit examples when $m>1$. Their existence is due
exactly to the same mechanism as in our previous lower dimensional
example. These solutions also correspond to wildly
Planck-scale-oscillatory configurations. In view of the result of the
previous section these regularizations correspond to bad suited
quantizations of the curvature part of the Hamiltonian constraint.

Unfortunately the construction of physical inner product of the
theory is not yet well understood and it is in this respect that
our argument cannot be as strong as the one made for 2+1 gravity
in the first part of this paper. Nevertheless, the fact that
quantizations of the theory in terms of $m>1$ produce these extra
local excitations, i.e. new degrees of freedom, strongly
discourages the choice of such theories. One should
expect these spurious solutions to be zero norm in the
physical inner product of loop quantum gravity.

\subsection{Quantization of the Hamiltonian constraint}

As explained in the introduction Thiemann's prescription leads to
$\hat H=\hat F(A) \widehat{ E E/{\rm det}(E)}$ as the only consistent
factor ordering in the quantization of the Hamiltonian constraint
(\ref{hamm}). We use the notation of reference  \cite{marcus}. With all
this in mind the action of the (regulated) quantum Hamiltonian
constraint on a spin-network vertex $\ket{v}$ is given by
\begin{equation}
  \label{H_m_Delta}
  \hat{\mathcal{H}}^m_{\Delta} \, \ket{v} = \frac{N_v i}{3 l_0^2 C(m)} \,
  \epsilon^{ijk} \, \mbox{Tr} \left[(
  {\hat{h}^{(m)}[\alpha_{ij}] - \hat{h}^{(m)}[\alpha_{ji}]}) \,
  \hat{h}^{(m)}[s_{k}]\, \hat{V} \,
  \hat{h}^{(m)}[s^{-1}_{k}] \right] \ket{v} ~,
\end{equation}
where the subindex $\Delta$ in $\hat{\mathcal{H}}^m_{\Delta}$ denotes
the triangulation used for the regularization of the action of the
constraint, $N_v$ is the value of the lapse function at the vertex,
and the supra-index $m$ denotes the fact that we are using the unitary
representation of spin $m$ to regularize the curvature term in terms
of the holonomies $\hat{h}^{(m)}[\alpha_{ij}]$ around to certain loops
$\alpha_{ij}$ and $\hat{h}^{(m)}[s_k]$ along segments $s_k$
respectively. The latter are defined in detail in \cite{marcus} and
will be graphically introduced in what follows.  The $m$-dependent
factor $C(m)$ is a normalization factor needed to satisfy the naive
continuum limit.

Now we will briefly remind the reader of the basic technicalities
associated to the quantization of the Hamiltonian constraint. In
this part of the paper we are following   \cite{marcus} almost
literally. For simplicity we use $3$-valent nodes in our pictures;
however, our argument is completely general and applies to
arbitrary $n$-valent nodes. We describe the regularization of
the Hamiltonian constraint by analyzing the action of the
different terms in (\ref{H_m_Delta}) separately. We start with the
action of the holonomy $\hat{h}^{(m)}[s^{-1}_{k}]$ operator on the
right which after a simple exercise of re-coupling theory gives
\begin{eqnarray}
  \label{holonomy_action1}
  \hat{h}^{(m)}[s^{-1}_{k}]~ \Bigg| \!\!
    \begin{array}{c}\mbox{\epsfig{file=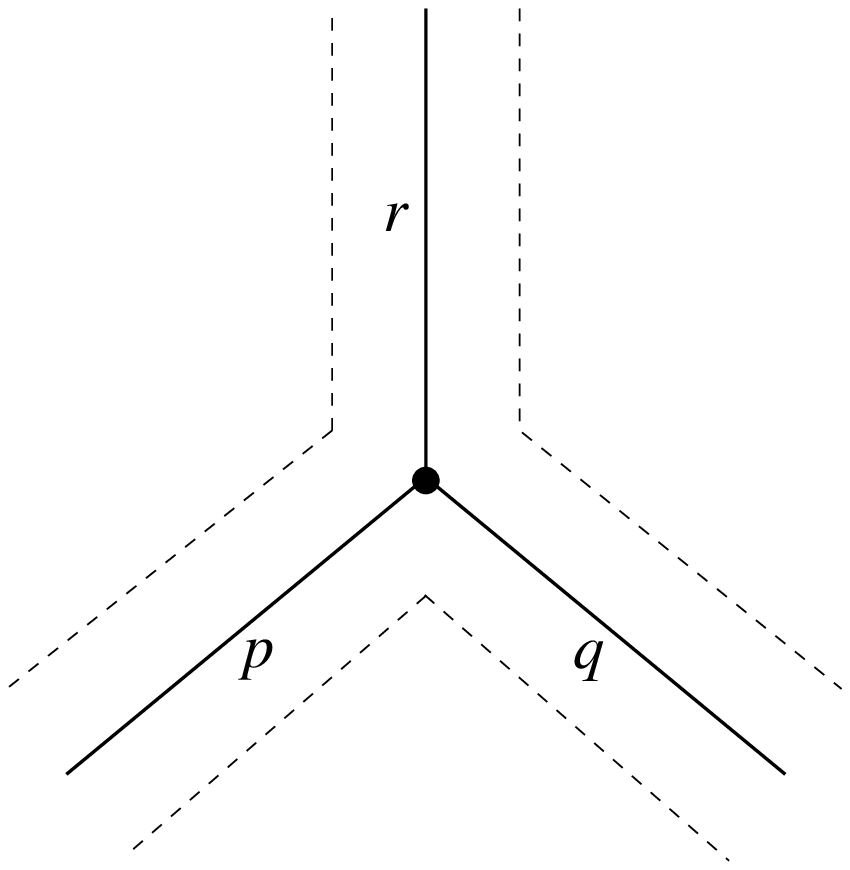,width=3.5cm}}
    \end{array} \!\!\Bigg\rangle
      ~= &\Bigg| \!
        \begin{array}{c}\mbox{\epsfig{file=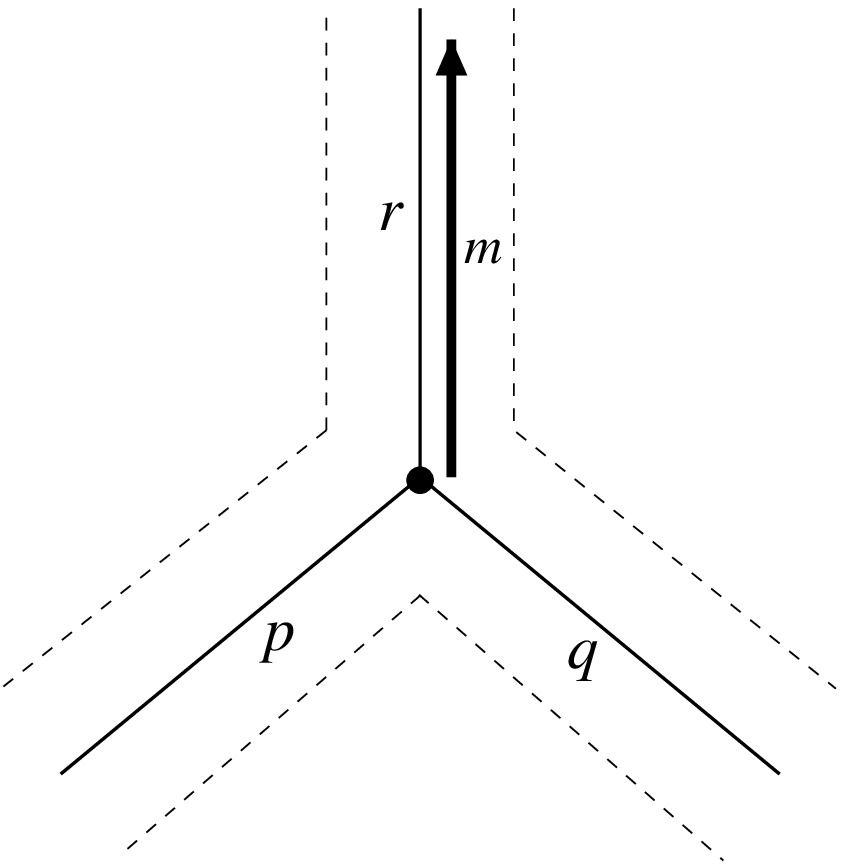,width=3.5cm}}
        \end{array} \!\!\Bigg\rangle& \\
      \label{holonomy_action2}
      = \;\;\sum_c
&\Bigg| \!\!
    \begin{array}{c}\mbox{\epsfig{file=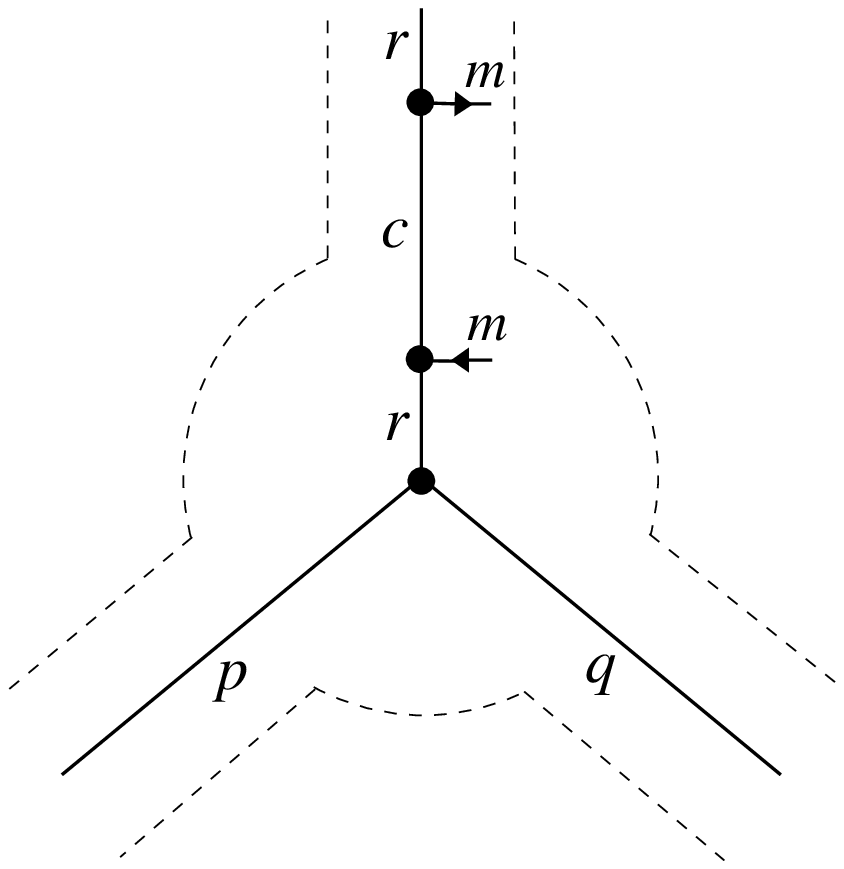,width=3.5cm}}
    \end{array} \!\!\Bigg\rangle& ~,
\end{eqnarray}
where the two new $3$-intertwiners are normalized. The
dotted line denotes a region of zero size introduced for
illustrative purposes. For instance, the vertical lines label by
representations $r$ and $m$ in the second diagram above are to be
thought of as overlapping.

The next operator appearing in (\ref{H_m_Delta}) from right to
left is the volume operator. Following the notation of
  \cite{marcus} the action of the volume operator on the vertex is given by
\begin{equation}
  \label{grapheqn_for_W}
  \hat{V}\hat{h}^{(m)}[s^{-1}_{k}]\, \ket{v}=\hat{V}
    \;\;\Bigg|\!\!
     \begin{array}{c}\setlength{\unitlength}{1.5 pt}
        \begin{picture}(50,40)
          \put(1,0){${\scriptstyle q}$}\put(1,34){${\scriptstyle p}$}
          \put(36,14){${\scriptstyle m}$}\put(46,34){${\scriptstyle c}$}
          \put(20,18){\line(-1,-1){15}}\put(20,18){\line(-1,1){15}}
          \put(30,18){\line(1,1){15}}\put(30,18){\line(1,-1){8}}
          \put(20,18){\line(1,0){10}}\put(24,21){${\scriptstyle \alpha}$}
          \put(20,18){\circle*{3}}\put(30,18){\circle*{3}}
          \bezier{20}(5,18)( 5,33)(25,33)
          \bezier{20}(45,18)(45,33)(25,33)
          \bezier{20}(5,18)(5,3)(25,3)
          \bezier{20}(45,18)(45,3)(25,3)
       \end{picture}
    \end{array} \Bigg\rangle
  = \sum_\beta V(p,q,m,c){}_\alpha{}^\beta
    \;\;\Bigg|\!\!
    \begin{array}{c}\setlength{\unitlength}{1.5 pt}
       \begin{picture}(50,40)
          \put(1,0){${\scriptstyle q}$}\put(1,34){${\scriptstyle p}$}
          \put(36,14){${\scriptstyle m}$}\put(46,34){${\scriptstyle c}$}
          \put(20,18){\line(-1,-1){15}}\put(20,18){\line(-1,1){15}}
          \put(30,18){\line(1,1){15}}\put(30,18){\line(1,-1){8}}
          \put(20,18){\line(1,0){10}}\put(24,21){${\scriptstyle \beta}$}
          \put(20,18){\circle*{3}}\put(30,18){\circle*{3}}
          \bezier{20}(5,18)( 5,33)(25,33)
          \bezier{20}(45,18)(45,33)(25,33)
          \bezier{20}(5,18)(5,3)(25,3)
          \bezier{20}(45,18)(45,3)(25,3)
       \end{picture}
    \end{array} \Bigg\rangle ~,
\end{equation}
where $V(p,q,m,c){}_\alpha{}^\beta$ denotes the matrix elements of
the volume operator, and the dotted region corresponds to a single
point. Inside this dotted region we graphically represent the
elements of the finite dimensional vector space ${\rm
Inv}[p\otimes q\otimes m \otimes c]$ in terms of normalized $3$-intertwiners (labelled by $\alpha$ and $\beta$ in the previous expression)
in the standard fashion. We recall that 3-valent nodes are used
here as a matter of convenience. In general the previous equation
remains true with the obvious modifications. As we will see below
our argument is completely independent of the volume-part of the
quantum Hamiltonian, and hence valid for any node valence. Next
one acts with the operators that represents the action of the
curvature tensor---the last term on the left of
(\ref{H_m_Delta})---obtaining
 \begin{eqnarray*}
   \label{hh_on_ng3vert}
   \lefteqn{({\hat{h}^{(m)}[\alpha_{ij}] -
        \hat{h}^{(m)}[\alpha_{ji}]}) \: \hat{h}^{(m)}[s_{k}] \!
   ~\;\Bigg|\!\!\!
   \begin{array}{c}\mbox{\epsfig{file=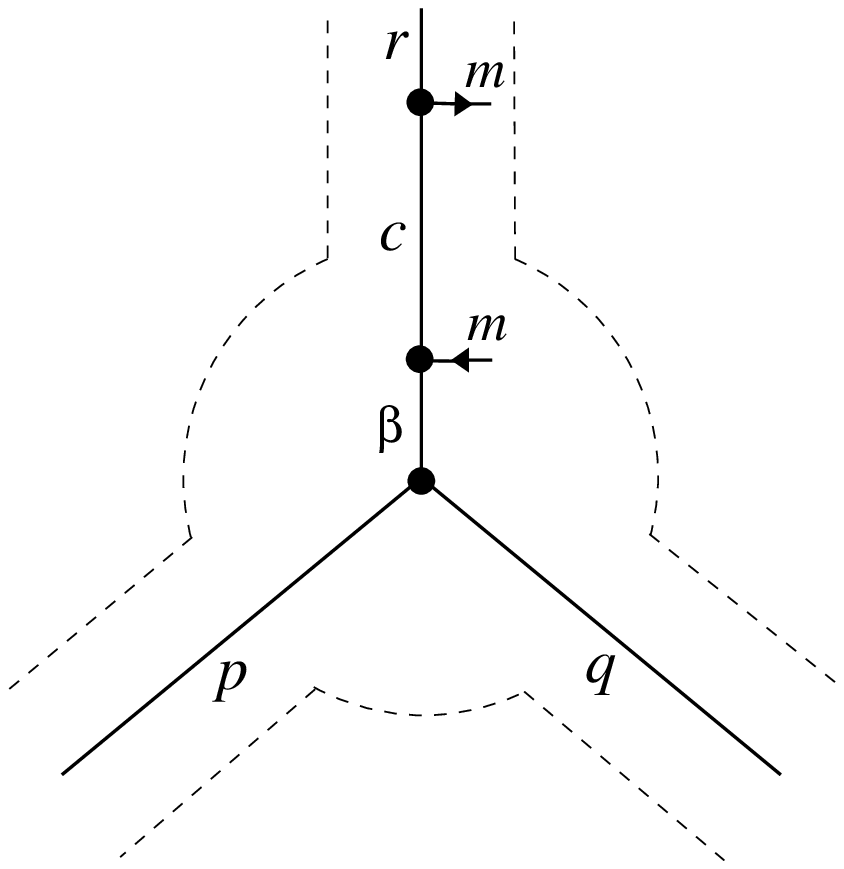,width=3.5cm}}
   \end{array} \!\!\!\Bigg\rangle }  \hspace{2cm} \\
   && =~ {(-1)^m} ~\left[\rule{0cm}{1.3cm}\right.
        ~\Bigg|\!\!\!
        \begin{array}{c}\mbox{\epsfig{file=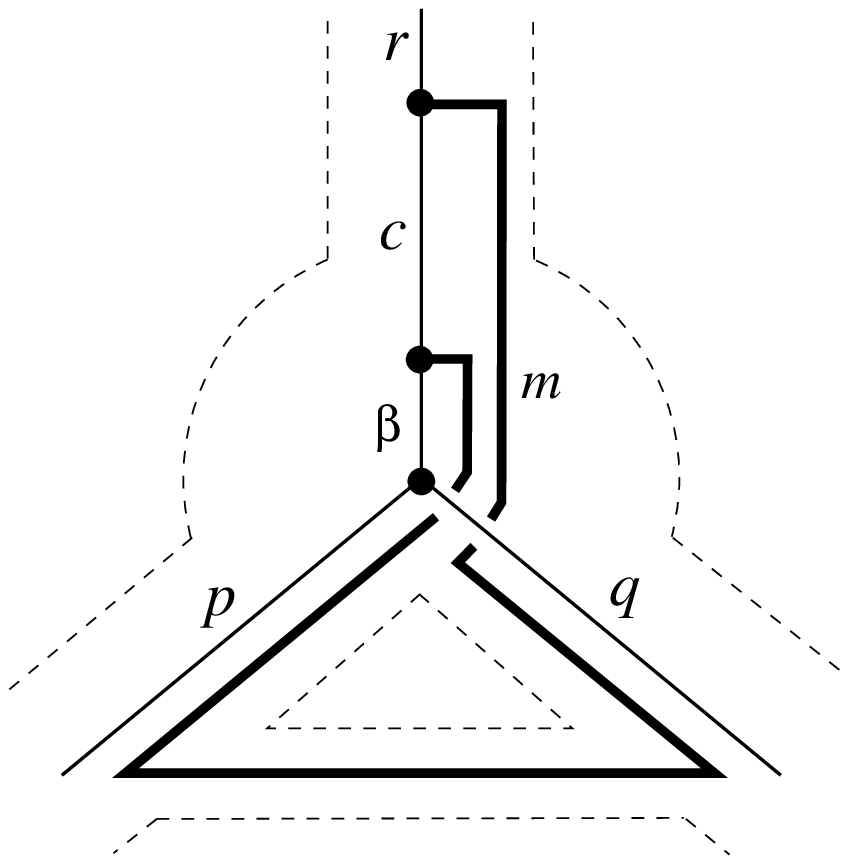,width=3.5cm}}
        \end{array} \!\!\!\Bigg\rangle
        ~-~ \Bigg|\!\!\!
        \begin{array}{c}\mbox{\epsfig{file=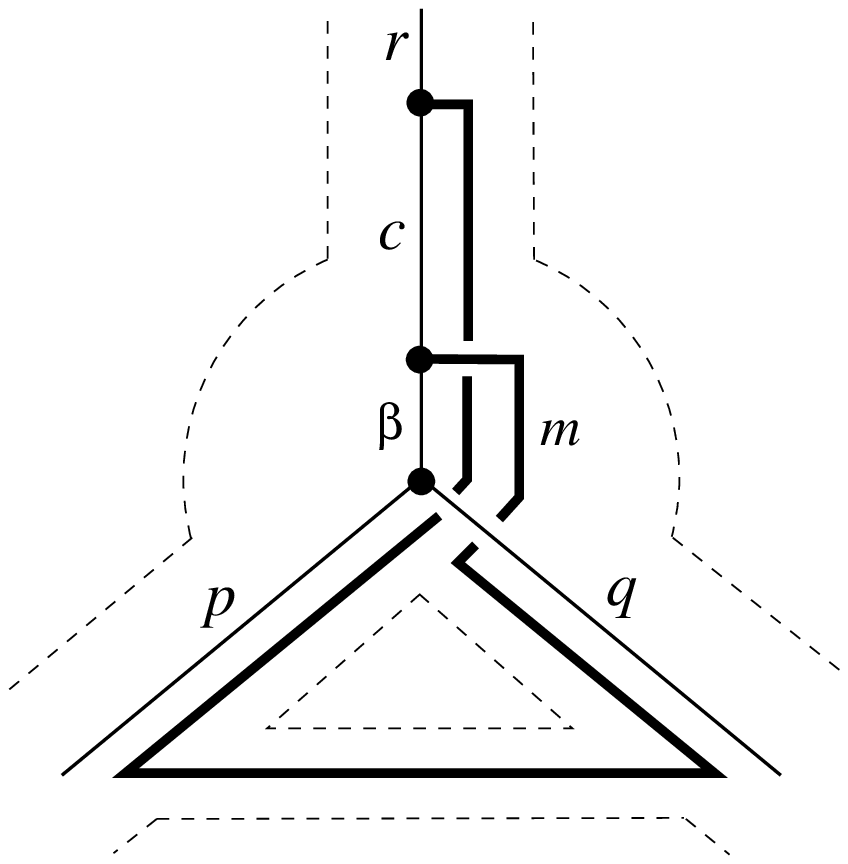,width=3.5cm}}
        \end{array} \!\!\!\Bigg\rangle \left.\rule{0cm}{1.3cm}\right]
         \hspace{1cm} \nonumber \\
\end{eqnarray*}
Putting all together and ignoring the pre-factor in
(\ref{H_m_Delta}) the action of the regulated Hamiltonian becomes
 \begin{eqnarray}
   \label{todo}
&& \nonumber \mbox{Tr}\left(
  ({\hat{h}^{(m)}[\alpha_{ij}] - \hat{h}^{(m)}[\alpha_{ji}]}) \:
  \hat{h}^{(m)}[s_{k}]\, \hat{V} \, \hat{h}^{(m)}[s^{-1}_{k}]
  \right) \Bigg|\! \begin{array}{c}\mbox{\epsfig{file=v1.eps,width=3.5cm}}
    \end{array} \!\!\Bigg\rangle
  \hspace{2cm} \\
   && =~{(-1)^m}\sum_{c \beta} V(p,q,m,c)^{\beta}_r
~\left[\rule{0cm}{1.3cm}\right.
        ~ \Bigg|\!\!\!
        \begin{array}{c}\mbox{\epsfig{file=v5.eps,width=3.5cm}}
        \end{array} \!\!\!\Bigg\rangle
        ~-~ \Bigg|\!\!\!
        \begin{array}{c}\mbox{\epsfig{file=v6.eps,width=3.5cm}}
        \end{array} \!\!\!\Bigg\rangle \left.\rule{0cm}{1.3cm}\right]
\end{eqnarray}
We call the new edge created by the action of the curvature {\em
exceptional edge}. This edge has special properties that grant the
absence of anomalies in the quantum theory (for details see
 \cite{bookt}).

Expanding the result in the spin network basis and projecting on
the connection representation we can write
\begin{eqnarray}
 \label{fin}
&& \lefteqn{
 <A|
\hat{\mathcal{H}}^m_{\Delta} \, |v (p,q,r) \rangle} \hspace{3.5cm}
=  \, \sum_{a,b}
        \, H^{(m)}(p,q,r; a,b)   ~\Bigg<A
\Bigg|
   \begin{array}{c}\mbox{\epsfig{file=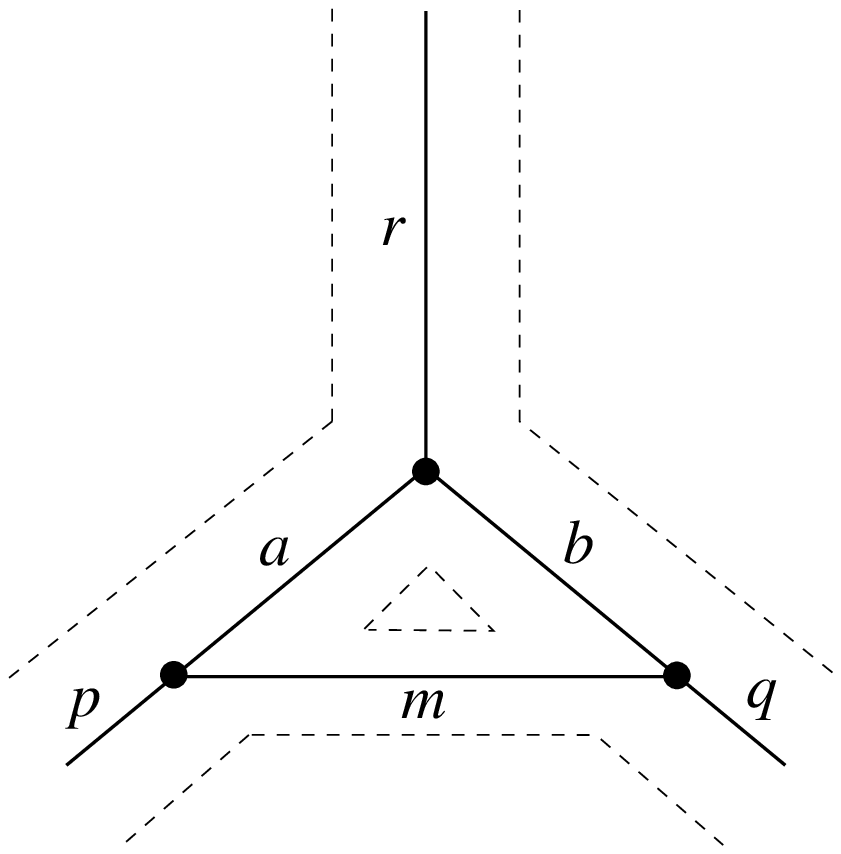,width=3.5cm}}
     \end{array}\!\!\!\Bigg\rangle + ...
= \nonumber \\
 & &=~ \sum_{a,b}
        \, H^{(m)}(p,q,r; a,b)\  \Psi^{p,q,r;a,b}(A_{out};A_{exc})
+...~,
\end{eqnarray}
where we have only explicitly written the term where the exceptional
edge is created on the bottom (there are two more terms in this case
but they are not important for the rest of the argument), and
$H^{(m)}(p,q,r; a,b)$ are the corresponding matrix elements of the
quantum Hamiltonian constraint. The functional
$\Psi^{p,q,r;a,b}(A_{out};A_{exc})$ is the spin network function of
the generalized connection along the edges of the underlying
graph. The variable $A_{exc}$ denotes the value of the holonomy along
the {\em exceptional} edge created by the action of the regulated
Hamiltonian constraint, which by appropriate gauge fixing at the
original vertex can be taken as the value of the holonomy around the
triangular loop created by the action of the constraint. On the other
hand $A_{out}$ denotes the generalized connection along the edges of
the spin-network graph which are different from the three edges
mentioned above.

It is important to notice that if we write $A_{exc}=x^{\mu}
\tau_{\mu}$, using the parametrization of $SU(2)$ of the previous
section, the action of the Hamiltonian constraint implies that
\begin{eqnarray}
 \label{parity}
 \sum_{a,b}
         H^{(m)}(p,q,r; a,b)  \Psi^{p,q,r;a,b}(A_{out};x^0,\vec x) = - \sum_{a,b}
        H^{(m)}(p,q,r; a,b)  \Psi^{p,q,r;a,b}(A_{out};x^0,-\vec x).
\end{eqnarray}
In other words the resulting state has a definite `parity' under
inversion of the generalized connection along the {\em
exceptional} loop as a consequence of equation (\ref{todo}). This
property will be important in the the following sections.

\subsection{Constructing solutions}\label{cstc}

We assume in this subsection that the corresponding vertex is $3$-valent. This will
simplify the discussion of the action of the quantum Hamiltonian
constraint.  This restriction is however a simple matter of
convenience as in that case the matrix elements of the quantum
constraint can be evaluated in a simpler way. In principle one
could generalize the argument presented here to arbitrary valence.
Notice however that such generalization is not necessary for the
validity of our conclusions as our objective is to show the
presence of spurious local degrees of freedom and not to fully
characterize them. In particular we will exhibit explicit spurious
solutions in the next subsection by means of a general argument
valid for arbitrary vertices.

We come back to equation (\ref{fin}) and the notation defined
there. Now we define a diffeomorphism invariant state
$(\Psi_{A_{out},x^{\mu}}|$ by \be\label{fati}
(\Psi_{A_{out},x^{\mu}}|=\sum_{\phi\in {\rm
Diff}(\Sigma)}\sum_{ab} \ {\Psi}^{p,q,r:a,b}(A_{out},x^{\mu})\
\Bigg<
   \begin{array}{c}\mbox{\epsfig{file=v10.eps,width=3.5cm}}
     \end{array}\Bigg|U_{\phi},
\ee where $U_{\phi}$ is the unitary operator that represents the
diffeomorphism $\phi$. The previous states are labelled by the
parameters $A_{out}$ and $x^{\mu}$ (or simply
$A_{exc}=x^{\mu}\tau_{\mu}$). The coefficients
${\Psi}^{p,q,r:a,b}(A_{out},x^{\mu})$ are given by the evaluation
of the corresponding spin network function defined in (\ref{fin})
for a definite choice of configuration, i.e., the generalized
connection (holonomies) along the edges of the corresponding
graph.

We also assume that the rest of the spin network state is
annihilated by the quantum Hamiltonian constraint acting on the
other vertices. This assumption is realized for example by a spin
network state that has no {\em exceptional edges} apart from the
one on the vertex of interest. More precisely, because the action
of the quantum Hamiltonian constraint creates exceptional links on
spin network states we have that $(\Psi|\hat H[N]\ s\!>=0$ if the
diffeomorphism invariant state $(\Psi|$ does not have any
exceptional edge. From this basic solution one can obtain
infinitely many solutions by adding local excitations---solutions
to the local conditions imposed by the Hamiltonian constraint at a
vertex---at different vertices. This is precisely what we do in
order to construct the new solution.

It is direct to check that for any spin-network state $|\phi>$ we have \be
(\Psi_{A_{out},x^{\mu}}|\hat H[N] \phi>= \left\{\begin{array}{ccc} 0\
\ \ \mbox{If the state $\phi\notin[\Psi-\mbox{exceptional edge}]$} \\
N_v P^{\va (2m)}_{A_{out}}(x^{\mu})\ \ \ \ \, {\rm otherwise}
\end{array}\right.,\ee
where $[\Psi-\mbox{exceptional edge}]$ denotes the equivalence class
under diffeomorphisms of $\Sigma$ of the spin network state obtained
form any element in the sum (\ref{fati}) by setting $m=0$, $a=p$ and $b=q$ respectively,
%
%
 and $N_v$ is
the value of the lapse function at the corresponding vertex.
The quantity $P^{\va (2m)}_{A_{out}}(x^{\mu})$
is an order $2m$ polynomial of the variable $x^{\mu}$ explicitly given by
\be
P^{\va (2m)}_{A_{out}}(x^{\mu})=\sum_{a,b}
        \, H^{(m)}(p,q,r; a,b)\  \Psi^{p,q,r;a,b}(A_{out};x^{\mu})
\ee

The coefficients of the previous polynomial can be shown to be
real: the reality of
$\Psi^{p,q,r;a,b}(A_{out};x^{\mu})$ follows from the fact that
spin network functions can be normalized to be real functions of
the generalized connection.  Spin networks can be taken as real
because they can be expressed as real linear combinations of
products of traces of Wilson loops in the fundamental
representation and hence real. The matrix elements of the
Hamiltonian constraint are also real in this basis. This might might seem
strange as the Hamiltonian constraint is not
self-adjoint. This is perhaps the reason why this property of the
Hamiltonian constraint has not been previously noticed in the
literature. It is a simple matter to proof the reality of the
matrix elements of the Hamiltonian for $3$-valent vertices
\footnote{Let us briefly support the statement of reality. In fact the
result is a simple consequence of properties of the reality properties
of the spin-network basis and the volume operator. From equation
(\ref{todo}) one concludes that the matrix elements of the quantum
Hamiltonian constraint are real if the matrix elements of the volume
operator are real (the reality of the combination of spin networks on
the right follows directly from the reality of spin network basis
elements).  Therefore it remains to show that the matrix elements of
the volume operator appearing in (\ref{todo}) are real. Recall that
the finite dimensional matrix $V(p,q,m,c){}_\alpha{}^\beta$ is defined
as $V=\sqrt{|W|}$ where $W$ essentially corresponds to the quantization
of $\epsilon_{abc}E^{a}_iE^{b}_jE^{c}_j\epsilon^{ijk}$. Acting on
finite valence nodes, and because its action does not change the
valence, $W$ can be represented by a finite dimensional hermitian
matrix. In order to define the square root one must go to the basis
that diagonalizes $W$, namely \be V=\sqrt{|W|}=U\sqrt{|W_D|} \ U^{-1}
\ee where $W_D$ is the diagonal form of $W$ \cite{marcus}. An important
property of $W$ is that it is purely imaginary and skew-symmetric
 \cite{DePietri:1996pj,Thiemann:1996au}. Hence $W^2$ is real and
symmetric $U$ is orthogonal from where it follows the reality of $V$.
This completes the proof of the reality of the matrix elements of the
quantum Hamiltonian constraint.}. It is not obvious whether the
reality holds for general matrix elements. This would be interesting
to explore.

The state $(\Psi_{A_{out},x^{\mu}}|$ would be in fact a physical state
for every solution $x^{\mu}$ of the equation $P^{\va
(2m)}_{A_{out}}(x^{\mu})=0$ with $x^{\mu}x_{\mu}=1$! As the order of the
polynomial increases with $m$, it is natural to expect that the number
of solutions of $P^{\va (2m)}_{A_{out}}(x^{\mu})=0$ will do so as
well. However, it could happen that for some reason non of the non
trivial solutions of the polynomial equation satisfy
$x^{\mu}x_{\mu}=1$.  Notice however that the reality of the
coefficients of $P^{\va (2m)}_{A_{out}}(x^{\mu})$ plus the fact its
coefficient depend on the external (continuum) parameters $A_{out}$
suggest that it should be possible to tune the polynomial equation so
that its solutions lay on the unit sphere. Nevertheless, in order to show this
explicitly one would need the explicit evaluation of the
matrix elements of the Hamiltonian constraint. This is not a serious obstacle
as such analysis for $3$-valent vertices would require a simple generalization of the
results of  \cite{Borissov:1997ji}. However, such a strategy will take us for a considerable
technical detour in the paper; so we will instead demonstrate the
existence of spurious solutions by a different method.

Assume for the moment that these solutions exist for $m>1/2$. The
existence of these solutions is directly linked to  our choice
of regularization indicating that the physically correct quantizations
must be those for which the curvature tensor is regularized in terms
of the fundamental representation. If on the contrary one wants to
insist in using a higher $m$ representation in the definition of the
theory one must provide a strong justification for the inclusion of
the extra local degrees of freedom. The understanding of the
construction of the physical inner product from the quantum
constraints would certainly make the result more robust. Our results
in 2+1 gravity suggest in this respect that the spurious solutions
appearing for higher $m$ regularizations would be of zero norm and
hence would disappear from $\Hp$.

\subsection{Solutions from an algebraic argument}\label{pilin}

Instead of explicitly computing the matrix elements of the quantum
Hamiltonian---which would present a quite formidable task---we
construct solutions in this section by a simple algebraic
argument. The idea is to make use of equation (\ref{parity}).
The argument presented here is valid for any vertex valence.

\subsubsection{Example in quantum mechanics}

As an example we consider a quantum mechanical particle on the unit
sphere. An orthonormal basis of the Hilbert space can be taken to be
the angular momentum basis whose elements we label $|\ell m>$
(s.t. $L^2|\ell m>=\ell(\ell+1)|\ell m>$ and $L_z|\ell
m>=m|\ell m>$). We have that the wave function
$<\vec x,\ell m>=Y_{\ell m}(\vec x)$  transforms under
parity as
\[Y_{\ell m}(-\vec x)=(-1)^{\ell} \, Y_{\ell m}(\vec x). \]
Due to the previous property the action of the parity operator $\hat p$
on basis elements is simply
\[\hat p \, |\ell m>=(-1)^{\ell} |\ell m>.\]
The action of our (toy) Hamiltonian constraint $\hat H$ is defined by
\be
<\vec x |\hat H \, |\ell m>=\sum_{n,q} h_{\ell m; n q} \   Y_{n q}(\vec x),
\ee
which is the simplified analog of equation (\ref{fin}). To complete
the analogy we require $\hat H$ to be such that $<\vec x |\hat H \,
|\ell m>=-<- \vec x |\hat H \, |\ell m>$ which can be achieved if $\hat
H=({1}-\hat p)\hat H_0$. In this analogy we associate
\[({1}-\hat p) \rightarrow\frac{\hat{h}^{(m)}[\alpha_{ij}] -
        \hat{h}^{(m)}[\alpha_{ji}]}{2}\] and
\[ \hat H_0\rightarrow\hat{h}^{(m)}[s_{k}]\, \hat{V} \, \hat{h}^{(m)}[s^{-1}_{k}]. \]

For an operator like this it is very easy to find solutions. In fact any dual state
of even parity will be obviously annihilated by $\hat H$. A basis of solutions will be
given by the states $<2n,m|$ for any positive integer $n$.

\subsubsection{Solutions of Thiemann's Hamiltonian}\label{sysy}

In order to find solutions of the quantum Hamiltonian we must
first construct states with a definite `parity' under the `reflection'
$A_{exc}\rightarrow A^{-1}_{exc}$. A family of candidate states are given by the following spin network states
\begin{eqnarray}
 \label{candidate}
&& {\Bigg<
   \begin{array}{c}\mbox{\epsfig{file=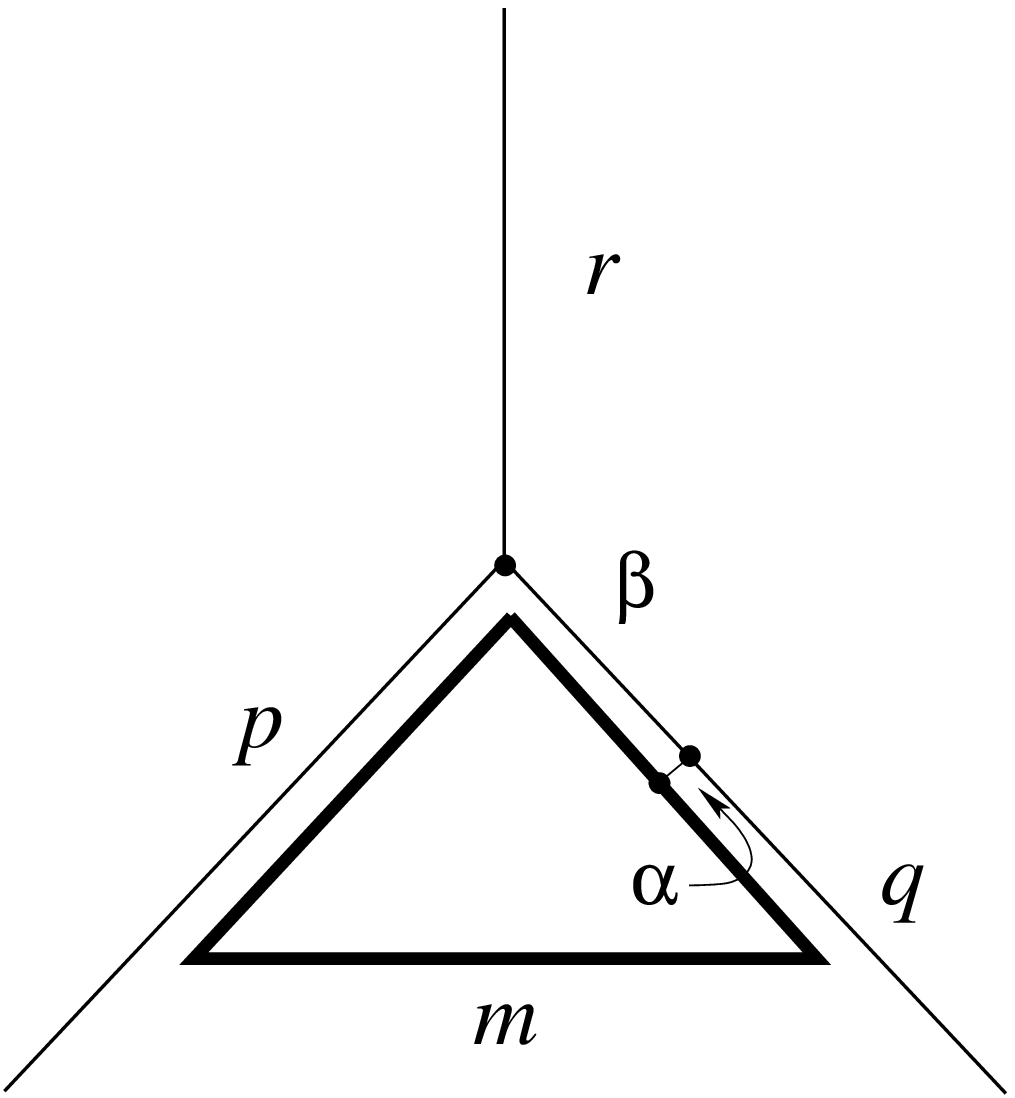,width=3cm}}
     \end{array}\!\!\!\Bigg|}
\end{eqnarray}
which under the transformation $A_{exc}\rightarrow A^{-1}_{exc}$
transform by a factor $(-1)^{\alpha}$   \cite{Borissov:1997ji}. The
next step is to find a diffeomorphism invariant state starting
from the previous spin network by means of summing over the action
of diffeomorphisms. The corresponding state is an element of the
set of distributions or linear functionals $Cyl^{\star}$ and can
be written as
\begin{eqnarray}
 \label{final} && { (\Psi|=\sum_{\phi \in {\rm Diff}[M]} \
 \sum_{\alpha \beta} \ c^{\va\Psi}_{\alpha \beta}\ \Bigg<
 \begin{array}{c}\mbox{\epsfig{file=soli.eps,width=3cm}}
 \end{array}\!\!\!\Bigg|\ {\rm U}[\phi]}
\end{eqnarray}
where $U[\phi]$ is the unitary operator that generates diffeomorphism
and the only condition on the coefficients is that $c^{\va \Psi}_{\alpha\beta}=0$
if $\alpha$ is odd.

Direct calculation shows that the previous is a solution
of the m-quantum Hamiltonian constraint, namely that $(\Psi,H^{\va (m)}[N]s>=0$
for any arbitrary $|s>\in \Hk$. The previous statement is non trivial
only in the  case when $|s>$ is in the diff-equivalent class of the spin network
state we started with. In that case the answer is zero because we are computing the
superposition between an even parity with an odd parity state which must vanish.

The solutions found in the previous subsection are labelled by two
quantum numbers $\alpha$ and $\beta$.  The set of possible values for
these two quantum numbers grows with the value of the ambiguity
parameter $m$. There are in fact $2m+1$ allowed values for $\alpha$
which lead to ${\rm IntegerPart}(m+1)$ even values.  If $m=1/2$ we
have only the possibility $\alpha=0$. However, if we use the
fundamental representation of $SO(3)$, i.e., $m=1$ we have two
possibilities: $\alpha=0$, already present in the previous case and
$\alpha=2$. This solution corresponds to a spin two local excitation!
For higher values of $m$ there are more solutions as an artifact of a
bad choice of regularization. According to the results in 2+1 gravity
these solutions should be regarded as spurious.

\subsection{Linear combinations }

One should also consider the
possibility of arbitrarily combining different $m$-regularizations
to produce a infinite-dimensional family of quantum Hamiltonian
constraints \be \label{hhaamm}\hat H[N]=\sum_m a_m \ \hat H_m[N] \ \ \
{\rm with} \ \ \ \sum_m a_m=1.\ee Now the previous solutions will
continue to exist since the action of the quantum constraint on them
is governed by a single term in the sum. The key equation is
\begin{eqnarray*}
        \Bigg(
        \begin{array}{c}\mbox{\epsfig{file=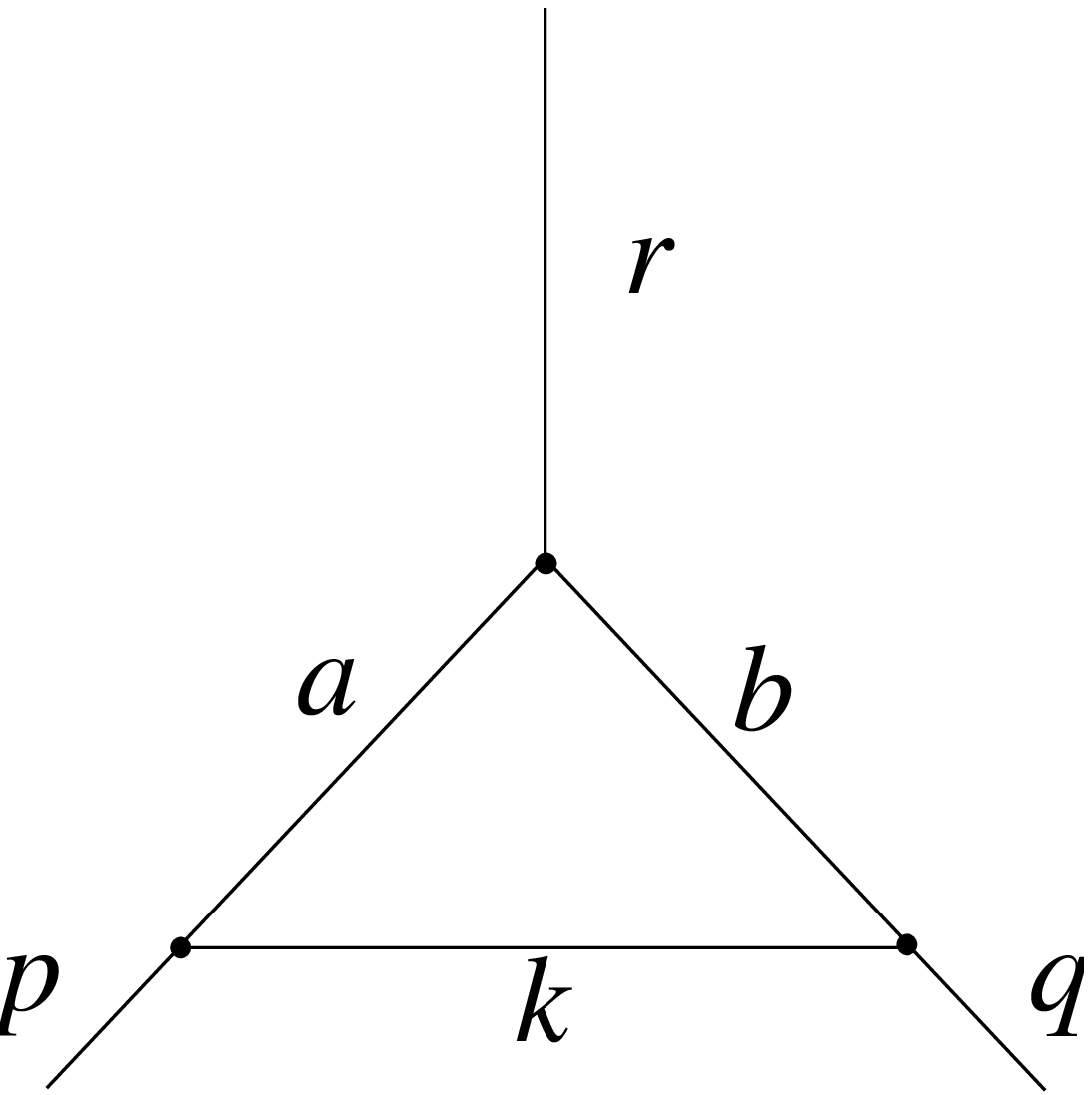,height=2cm}}
        \end{array},
        ~\sum_m a_m \ \hat H_m[N]~ \!\!\!
        \begin{array}{c}\mbox{\epsfig{file=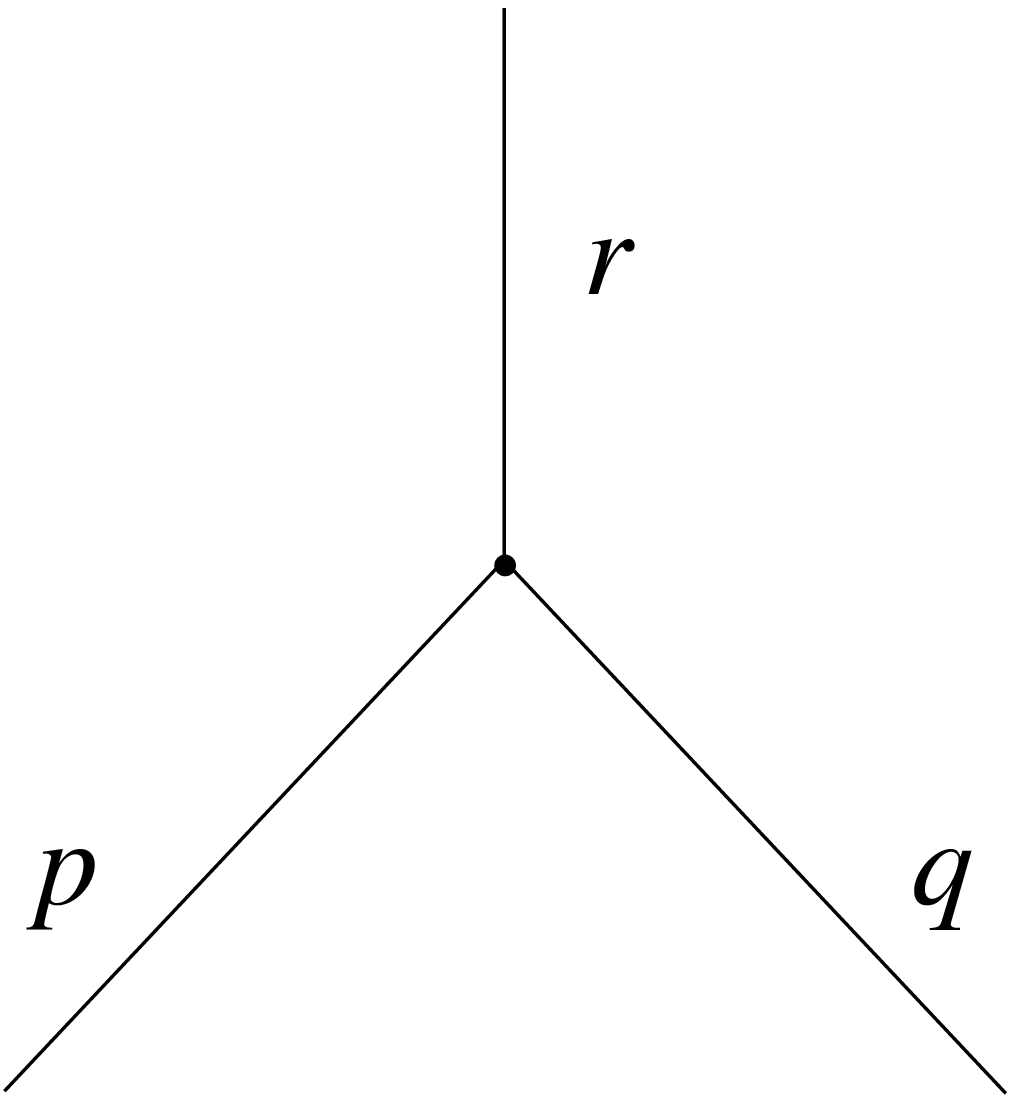,height=2cm}}
        \end{array} \!\!\!\Bigg\rangle=\Bigg(
        \begin{array}{c}\mbox{\epsfig{file=confi1.eps,height=2cm}}
        \end{array},
        ~\hat H_k[N]~ \!\!\!
        \begin{array}{c}\mbox{\epsfig{file=confi.eps,height=2cm}}
        \end{array} \!\!\!\Bigg\rangle,
\end{eqnarray*}
where the $(s|$ denotes a diff-invariant state associated to the spin
network $<s|$.  The validity of the previous equation allows for the
construction of spurious solutions by simply using the spurious
solutions found in the previous section for individual terms in
(\ref{hhaamm}). This seems quite different from what we found in
Section~\ref{lc}, where some linear combinations would lead to
quantizations that were equivalent to the one based on the fundamental
representation.

Even though this might be interpreted as a positive result one
should keep in mind that this happens because of a property of
Thiemann's quantum constraint that is also seen as a problem. More
precisely, the fact that among the solutions of Thiemann
Hamiltonian constraint there is a vast set of solutions of a
rather trivial nature. For example a diffeomorphism invariant
state labelled by a spin network with no exceptional edge is a
trivial solution of the constraints. 
This is related to the special character of the
exceptional edges that are added by the action of $H$ required by
the conditions that imply the absence of an anomaly
 \cite{Thiemann:1996aw}. The triviality of these solutions is
puzzling and seem to indicate that the restrictions imposed by
quantum constraint quantized {\em \`a la Thiemann} are too week to
lead to a theory with propagating degrees of freedom
 \cite{Smolin:1996fz}.  This problem is one of the main motivation
for the exploration of alternative definitions of the dynamics
such as the one proposed in the master constraint program, the
consistent discretization approach and the covariant spin foams
approach.


\section{Discussion}

The absence of divergences in the quantization of the Hamiltonian
constraint is a remarkable feature of loop quantum gravity. In
this work we point out that this important characteristic of the
theory does not, by itself, resolves the issue of renormalization
in quantum gravity as having a sound mathematical framework (free
of infinities) is intimately related to the existence of
ambiguities. In the case of loop quantum gravity there is a
infinite dimensional space of possible theories. Until the problem
of the ambiguities is settled the situation, regarding the
predictive power of the theory, is not much different from that of
standard perturbative approaches. In this paper we investigated
the so-called {\em $m$-ambiguity} associated to the unitary
representation used in the quantization of the configuration
variables. In the case of $2+1$ gravity the problem is completely
resolved. In $3+1$ gravity we provide evidence pointing at a
possible resolution of the question. In what follows we discuss
these results in more detail.

\subsubsection*{2+1 loop quantum gravity}

We have showed that consistency of the quantum theory eliminates
the ambiguities related to the quantization of the curvature
constraint in $2+1$ loop quantum gravity. If the regularization is
not performed using the holonomy in the fundamental representation
of the gauge group the appearance of extra (spurious) solutions
conspire against the possibility of removing the regulator in the
definition of the physical scalar product. There are other
prescriptions that lead to a well defined theory but they are
fully equivalent to the quantum theory defined in terms of the
fundamental representation. Pure gravity in three dimensions is an
example of theory belonging to the first class mentioned in the
introduction.

The spurious solutions to the quantum constraint regulated with
the representation $m$ (with $m>1/2$) correspond to wildly
oscillatory curvature configurations down to the Planck scale.
These solutions are annihilated by the regulated constraint but
because of the latter feature they are not well defined in the
`continuum' (i.e., independently of the regulator). Nevertheless,
if one defines the physical inner product in terms of the good
regularizations (e.g., $m=1/2$) then the spurious solutions of the
regulated constraint for the bad quantizations (e.g., $m>1/2$)
have zero physical norm.

Because $2+1$ gravity is a topological theory, the fact that the issue
of ambiguities can be completely settled in this case is not entirely
surprising---the renormalizability of $2+1$ gravity is advocated since
Witten's seminal work  \cite{Witten:1988hc}.  Gravity in $2+1$
dimensions has finitely many degrees of freedom and from this
perspective one would not expect serious difficulties dealing with the
UV problem. Our results make the previous statement precise in the
framework of loop quantum gravity and provides the starting point for
the analysis of the issue in $3+1$ dimensions. The results of the
first part of this work extends trivially to the case of spinning
particles coupled to 2+1 gravity studied in  \cite{a21}.

\subsubsection*{3+1 loop quantum gravity}

The effects of the {\em $m$-ambiguity} in 3+1 loop quantum gravity are
similar. Regularizing the holonomies used in the quantization of the
Hamiltonian with unitary representations of spin $m>1$ introduces new
local degrees of freedom. These solutions correspond, as in the lower
dimensional case, to highly oscillatory excitations at the Planck
scale. The mechanism leading to the existence of such solutions is the
analog of the 2+1 case: higher representation regularizations of the
curvature tensor appearing in the Hamiltonian constraint correspond to
functions on the groups with additional roots.

The direct computation of the spurious solutions of
Section~\ref{cstc} would require the explicit computation of the
matrix elements of the Hamiltonian constraint for arbitrary
regularizations. In Section~\ref{sysy} we used a symmetry argument
to explicitly exhibit the existence of new local degrees of
freedom associated with the choice of higher $m$ quantizations.
These local degrees of freedom correspond to higher spin local
excitations---for example the quantum number $\alpha$ in equation
(\ref{final}) takes values $\alpha=4,\cdots,2m$ for
$m=\mbox{integer}$.

At this stage one cannot construct a complete argument as in 2+1
gravity due to the lack of an explicit definition of the physical
inner product in 3+1 gravity. More precisely we cannot prove that the
spurious solutions would spoil the existence of a well defined
continuum limit unless they are zero norm in $\Hp$. Nonetheless the
existence of spurious solutions of the quantum constraints associated
to $m>1$ quantizations provides an argument against such theories that
changes our perspective regarding the ambiguity problem: if one would
like to use values of $m>1$ in the quantization one would need to
provide a clear justification for the inclusion of the associated
extra degrees of freedom. This is evidence pointing in the right
direction, we hope that future studies could shed more light on this
important issue.

Finally, let us mention that a study of the effects of the {\em $m$-ambiguity}
in the quantum mechanical context of loop quantum cosmology has been
performed in  \cite{Vandersloot:2005kh}. The results are consistent
with the ones presented here for the field theory. In fact there are
new solutions associated to a higher $m$-quantizations the Hamiltonian
constraint.  Most of these solutions are un-physical or
spurious in view of certain semiclassicality criteria
 \cite{Bojowald:2001xa} applied in the context of loop quantum
cosmology. It is interesting to notice that in the simple model
studied in  \cite{Noui:2004gy} one can also study the effects of the ambiguity
with the advantage of knowing the physical inner product. In this case
one can explicitly show that spurious solutions are indeed zero-norm
states. This is an interesting result pointing into an encouraging
direction. Yet one must keep in mind that this is a toy model which
lacks local degrees of freedom. Moreover, even though spurious
solutions are not in $\Hp$, the properties of the physical states do
depend on the ambiguity.

\subsubsection*{Physical Hamiltonian and other approaches to dynamics}

Our analysis in 3+1 gravity has been performed entirely in the
context of the framework of Thiemann's quantization of the Hamiltonian
constraint. Even when Thiemman's prescription provides a
mathematically consistent quantum operator, concerns have been raised
about its physical viability. Problems related to the so-called ultra-local
character of the quantum dynamics---which are rooted in the way
the constraint algebra of gravity is represented (for a review see
 \cite{Nicolai:2005mc})---have been pointed out as a
serious obstacle for the theory to reproduce general relativity in the
classical limit  \cite{Smolin:1996fz} (for a different perspective of
the same problem see
 \cite{Perez:2004hj}).

This has motivated the search for alternative definition of
dynamics such as: the covariant definition given by the so-called
{\em spin foam models}  \cite{a18}, alternative quantizations
proposed by Thiemann in his {\em master constraint program}
 \cite{Thiemann:2003zv}, and the program of Gambini and Pullin of
{\em consistent discretizations}  \cite{Gambini:2004vz}. In the latter
two alternative formulations similar regularization problems give raise
to ambiguities which are the analog of the $m$-ambiguity studied here.
Therefore, the questions raised by this article must also be addressed
in these cases.

Since our argument is based on the existence of multiple solutions
of the quantized constraints we expect its conclusions to be
sufficiently general to provide a non trivial insight in cases in
which the details of the dynamics are different. In Fact, in the
first part of the paper we showed how the analysis of the
ambiguity in the canonical formulation of 2+1 gravity had a
precise parallel in the covariant formulation (or spin foam
representation) of the theory. For this reason we think that our
results obtained in the context of Thiemann's constraint should
apply in suitable form to any definition of the quantum dynamics
where the connection is represented by holonomies.

\subsubsection*{Spin foam models from constrained BF theory}

In Section~\ref{sfm} we showed how the potential ambiguities
arising in the definition of the path integral of BF theory can be
eliminated. Our results in three dimensions can be easily
generalized to arbitrary dimensions. Therefore, there are no
ambiguities of the type analyzed here in the quantization of BF
theory in four dimensions. This provides extra incentive for the
search of a covariant formulation (or spin foam representation)
based on the idea of viewing gravity as a constrained BF theory.
Many of the spin foam models studied in the literature are of this
kind  \cite{baez5,ori2}. Particularly attractive in this respect is
the treatment proposed by Freidel and Starodubtsev
 \cite{Freidel:2005ak}.

\subsubsection*{General considerations about first order gravity}

In the introduction we advocated the similarities between the
renormalization problem in perturbative and loop quantum gravity
with the purpose of stressing the importance of a clear
understanding of the ambiguity issue in the latter. Now we would
like to point out an important difference which provides an
independent (heuristic) argument supporting the idea the
background independent quantum field theory of gravity pursued by
loop quantum gravity should be rather restrictive instead of
infinitely ambiguous.

Loop quantum gravity---or spin foam models as their covariant
formulation---is a general framework for the non perturbative
quantization of gravity in the first order formulation. By the
first order formulation we mean here the most general
diffeomorphism invariant theory that one can write in terms of a
tetrad of 1-forms and a Lorentz connection $A$ \footnote{The
canonical quantization of these theories directly leads to the
fundamental variables of LQG: fluxes of non-Abelian electric field
and generalized connections.}. The most general form of such
action in three dimensions is \be \label{reno} S[e,A]=\int \ {\rm
Tr}[e\wedge F(A)] +\Lambda \int {\rm Tr}[e\wedge e\wedge
e\epsilon], \ee which was first quantized and argued to be
renormalizable by Witten  \cite{Witten:1988hc}. In four dimensions
the most general action becomes \ba\nonumber &&\!\!\!\!
S[e,A]=\frac{1}{2\kappa}\int {\rm Tr}[e\wedge e\wedge
F^{\star}(A)] +\frac{1}{\kappa\gamma}\int {\rm Tr}[e\wedge e\wedge
F(A)]+\\  &&\ \ \ \ \ \ \  + \Lambda \ \int {\rm
Tr}[e\wedge e\wedge (e\wedge e)^{\star}] + \alpha \int {\rm
Tr}[F(A)\wedge F^{\star}(A)]+\beta \int {\rm Tr}[F(A) \wedge
F(A)],\label{sisi} \ea where $\gamma$ is the Immirzi parameter,
and $\alpha$ and $\beta$ are coupling constants. Notice that from
this perspective it is natural to introduce a non trivial Immirzi
parameter which is essential for the definition of loop quantum
gravity.

Heuristically, in standard renormalization framework, the
simplicity of the previous action is reminiscent of a
`renormalizable' theory: all the possible terms compatible with
the postulated fundamental symmetries are finitely many
\footnote{Here we are assuming that there are no matter couplings.
In order to couple the theory to standard matter one need to use
the inverse tetrad $e^{-1}$ which is not a fundamental variable.
Notice that fermions can be brought into the game without
introducing the inverse tetrad.}. However this argument cannot be
made in the standard way because the previous action is not
quadratic around the diffeomorphism invariant vacuum $e=0$ and
$A=0$ and one cannot make use the usual perturbative treatment
 \cite{Witten:1988hc}. Moreover, if one instead defines the
perturbative theory around an invertible configuration, say
$e_a^I=\delta_a^I$, then the perturbation theory generates the
infinitely many terms in the effective action that can be written
in terms of the inverse $e^{-1}$. Hence we arrive in this case to
the standard {\em no-renormalizability} of gravity. If the
striking simplicity of the general action (\ref{sisi}) is of an
indication in some sense of the uniqueness of the associated
quantum theory the question must be explored non-perturbatively.

Loop quantum gravity and spin foam models are non-perturbative
approaches based on this action. The fundamental excitations,
spin network states, represent in fact quantum geometries that are
degenerate almost everywhere. Indeed, strictly speaking states
corresponding to non degenerate geometries do not
exist. Only complicated superpositions of polymer-like excitations
approximate metric configurations such as $e_a^I=\delta_a^I$ in
the weak sense given by {\em coarse graining}  \cite{Bombelli:2004si,Ashtekar:1992tm}: probing
the state at low energies yields a metric manifold while the
geometry is almost-everywhere degenerate ($e=0$) at the Plank
scale. The simple form of the action (\ref{sisi}) in terms of
these variables suggests that the resulting quantum theory could be
rather restrictive.

If there are no ambiguities at the fundamental level, then how is one
to recover the infinite series of higher dimensional operators in the
effective action of gravity? Coarse graining would be the
mechanism. In the semi-classical limit the quantum geometry states
approximate a space-time geometry when probed at sufficiently low
energies. Deviations from the classical behavior due to quantum
fluctuations will appear as higher powers of the curvature tensor
corrections in the effective action because $e^{-1}$ now exists in the
coarse-grained sense. In this process only coarse graining would
generate the higher curvature corrections in the effective action
description. These terms should be calculable from the fundamental
theory and the properties of the semiclassical states considered. In
other words, the non-perturbative formulation of first order gravity
(where ambiguities are controlled by a finite number of parameters)
could play the role of `{\em renormalizable}' theory underlying the
non renormalizable metric gravity. From this perpective we could
expect that, as in 2+1 gravity, the (infinite dimensional set of)
regularization ambiguities in the quantization would have to be
drastically reduced in the definition of$\Hp$. This question will have
to be explored further in future work; our present results
provide some supporting evidence in this direction.

\subsubsection*{Gravitons?}

\begin{figure}\label{graviton} \centerline{\hspace{0.5cm} \(
\begin{array}{c}
\includegraphics[height=3cm]{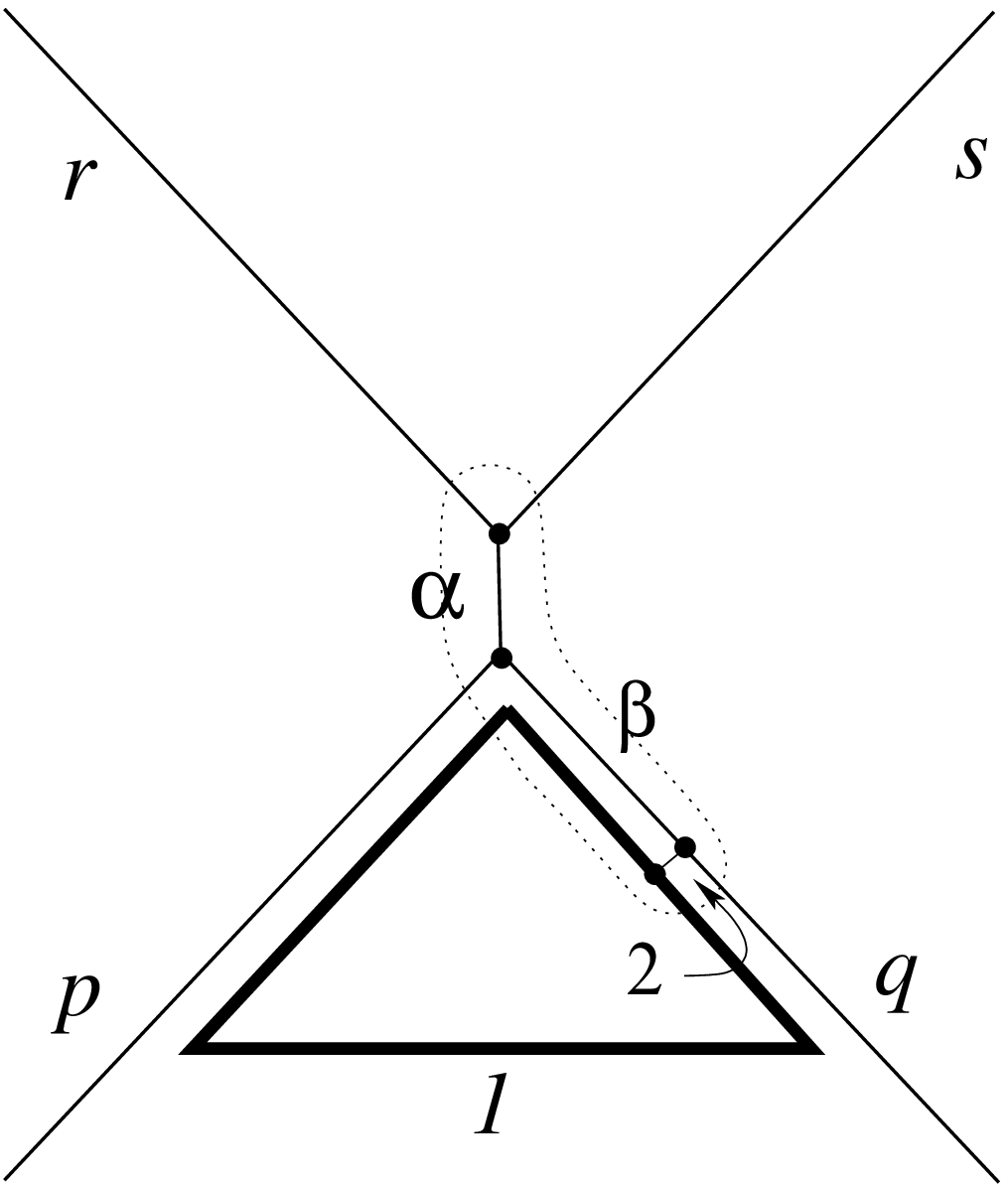}
\end{array}\ \ \ \ \ \rightarrow \ \ \ \ \ \ \begin{array}{c}
\includegraphics[height=3cm]{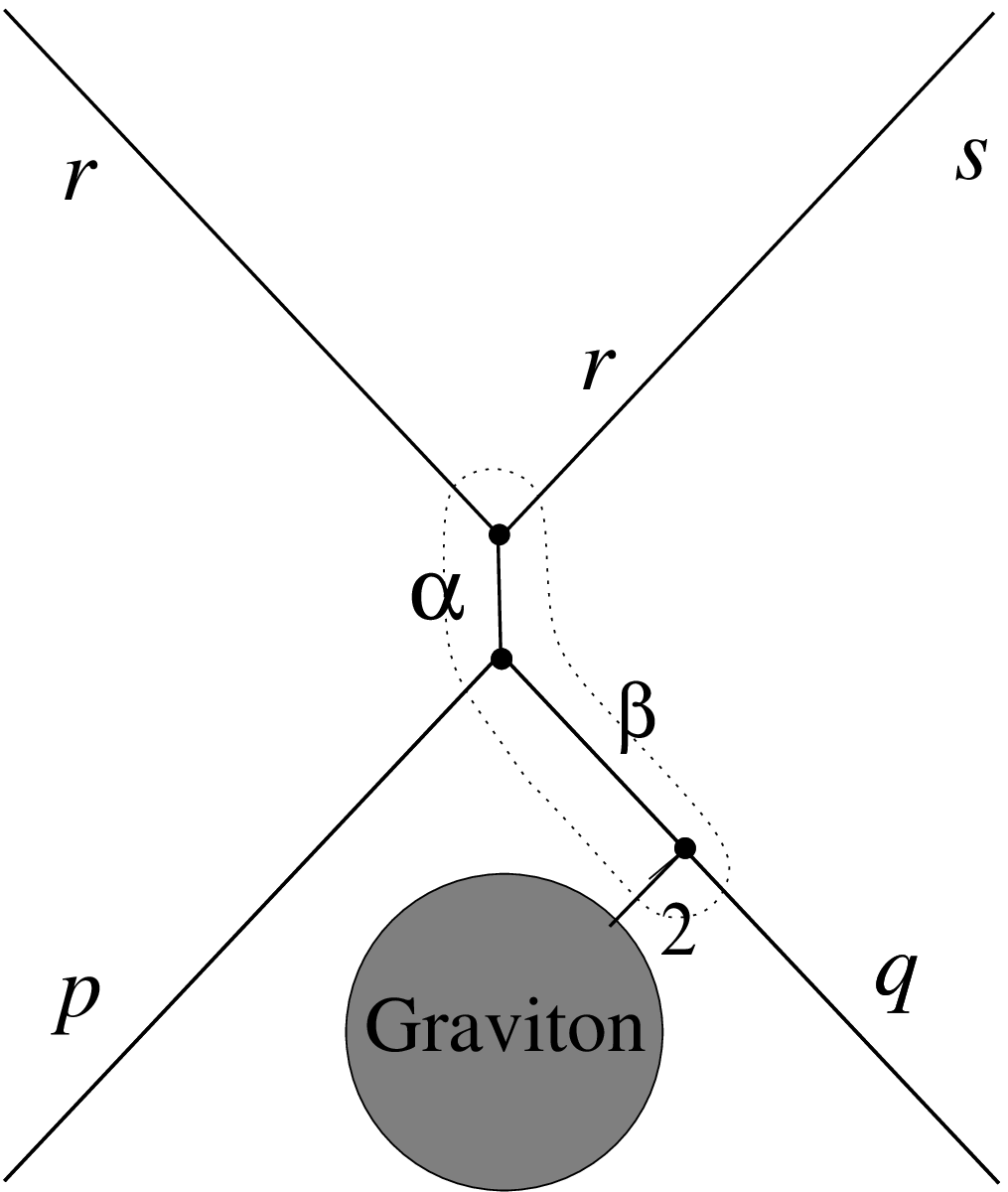}
\end{array}\)}
\caption{Interpretation of the solutions (\ref{final}) for $m=1$ as
graviton excitations. Starting from a solution to the
constraints given by a diff-invariant spin network with a vertex with no exceptional
edge we can construct a new solution as explained in Section \ref{pilin}
and illustrated here. The solution space is parametrized by the
quantum numbers $\alpha$ and $\beta$ in this figure. The dotted region
corresponds to a single point in the spin network graph.}
\end{figure}

Let us conclude our discussion with a speculative interpretation
of an intriguing type of solution to the quantum Hamiltonian
constraint found in Section~\ref{pilin}. We constructed an
argument to rule out higher spin regularizations of the quantum
Hamiltonian. The case $m=1/2$ and $m=1$ are special as they
correspond to the fundamental representation of $SU(2)$ and
$SO(3)$ respectively. Therefore, we might expect the quantization
based on $m=1$ to be of interest.  In this case the solutions
found in Section~\ref{pilin} have a clear-cut interpretation as
spin two excitations. It would be interesting to further
investigate the possibility of interpreting the solutions
presented in (\ref{final}) as the fundamental degrees of freedom
leading, in the low energy limit, to the notion of graviton.
Notice that if we assume that the continuum limit is dominated by
four valent vertices (i.e., {\em quantum tetrahedra:} the simplest
excitations of $3$-volume), these solutions are labelled by two
local quantum numbers as illustrated in Figure \ref{graviton}.  In
this speculative interpretation we see the infinitesimal loop that
is attached to the geometry by a link labelled with $\alpha=2$ as
a spin two particle. Notice that this is fully analogous to the
way in which spin $1/2$ fermions are coupled to the geometry
 \cite{baez11,c10bis}. 

\section{Acknowledgement}

The author would like to thank F. Conrady, S. Lazarini, D. Marolf,
D. Perini, C. Rovelli, D. Sudarsky, and K. Vandersloot for discussions and
suggestions.  The idea of this work was motivated by questions of R.
Wald at the VI Mexican School on Gravitation and Mathematical
Physics.


\end{document}